\documentclass{aa}  

\usepackage{graphicx,xtab}
\usepackage{siunitx}
\usepackage{txfonts}
\usepackage{multirow}
\usepackage{xcolor}
\usepackage{multicol}

\usepackage[]{hyperref}
\hypersetup{colorlinks = true,citecolor = blue}
\newcommand{\rr}{\mathcal{R}}
   
\begin{document}

   \title{LoTSS/HETDEX: Optical quasars\\ \Large{I. Low-frequency radio properties of optically selected quasars}}

   \author{G\"ulay G\"urkan
          \inst{1*},
          M. J. Hardcastle\inst{2}, P. N. Best\inst{3}, L. K. Morabito\inst{4}, I. Prandoni\inst{5}, M. J. Jarvis\inst{4,6}, K. J. Duncan\inst{7}, G. Calistro Rivera\inst{7}, J. R. Callingham\inst{8}, R. K. Cochrane\inst{3}, J.H. Croston\inst{9}, G. Heald\inst{1}, B. Mingo\inst{9}, S. Mooney\inst{10}, J. Sabater\inst{3}, H. J. A. Röttgering\inst{7}, T. W. Shimwell\inst{8}, D. J. B. Smith\inst{2}, C. Tasse\inst{11,12} and W. L. Williams\inst{2}}

   \institute{CSIRO Astronomy and Space Science, PO Box 1130, Bentley WA 6102, Australia\\
              \email{gulay.gurkan.g@gmail.com}
              \and
             Centre for Astrophyics Research, School of Physics, Astronomy and Mathematics, University of Hertfordshire, College Lane, Hatfield AL10 9AB, UK
             \and
             SUPA, Institute for Astronomy, Royal Observatory, Blackford Hill, Edinburgh, EH9 3HJ, UK
             \and
             Astrophysics, University of Oxford, Denys Wilkinson Building, Keble Road, Oxford OX1 3RH, UK      
             \and
             INAF - Istituto di Radioastronomia, Via P. Gobetti 101, 40129 Bologna, Italy
             \and
             Physics and Astronomy Department, University of the Western Cape, Bellville 7535, South Africa
             \and
             Leiden Observatory, Leiden University, PO Box 9513, NL-2300 RA Leiden, the Netherlands
             \and
             ASTRON, the Netherlands Institute for Radio Astronomy, Postbus 2, 7990 AA Dwingeloo, The Netherlands
             \and
             School of Physical Sciences, The Open University, Walton Hall, Milton Keynes, MK7 6AA, UK
             \and
             School of Physics, University College Dublin, Belfield, Dublin 4, Republic of Ireland
             \and
             GEPI, Observatoire de Paris, CNRS, Universite Paris Diderot, 5 place Jules Janssen, 92190 Meudon, France
             \and
             Department of Physics \& Electronics, Rhodes University, PO Box 94, Grahamstown, 6140, South Africa}

   \date{Received 18 July 2018 / Accepted 10 October 2018}

  \abstract{The radio-loud/radio-quiet (RL/RQ) dichotomy in quasars is still an open question. Although it is thought that accretion onto supermassive black holes in the centre the host galaxies of quasars is responsible for some radio continuum emission, there is still a debate as to whether star formation or active galactic nuclei (AGN) activity dominate the radio continuum luminosity. To date, radio emission in quasars has been investigated almost exclusively using high-frequency observations in which the Doppler boosting might have an important effect on the measured radio luminosity, whereas extended structures, best observed at low radio frequencies, are not affected by the Doppler enhancement. We used a sample of quasars selected by their optical spectra in conjunction with sensitive and high-resolution low-frequency radio data provided by the LOw Frequency ARray (LOFAR) as part of the LOFAR Two-Metre Sky Survey (LoTSS) to investigate their radio properties using the radio loudness parameter ($\rr = \frac{L_{\mathrm{144-MHz}}}{L_{\mathrm{i\,band}}}$). The examination of the radio continuum emission and RL/RQ dichotomy in quasars exhibits that quasars show a wide continuum of radio properties (i.e. no clear bimodality in the distribution of $\rr$). Radio continuum emission at low frequencies in low-luminosity quasars is consistent with being dominated by star formation. We see a significant albeit weak dependency of $\rr$ on the source nuclear parameters. For the first time, we are able to resolve radio morphologies of a considerable number of quasars. All these crucial results highlight the impact of the deep and high-resolution low-frequency radio surveys that foreshadow the compelling science cases for the Square Kilometre Array (SKA).}
   \keywords{quasars: normal -- optical:quasars -- radio:quasars}
  
  \titlerunning{Quasars in the LOTSS/HETDEX field}
  \authorrunning{G\"urkan et al.}
  \maketitle

%

\section{Introduction}

\label{sec:intro}

The radiative and jet power in active galactic nuclei (AGN) is generated by accretion of material on to massive galactic-centre black holes \citep{1995wref64,1998wref65,2000wref66}. However, more than one accretion mechanism is needed to explain the observed properties of the whole AGN population \citep[e.g.][]{1994intr55,2000intr61,2015intr131}, and the relationship between the radio emission, often generated by the interaction of a jet with its environment, and the radiative power generated by the accretion disc is complex. In quasars (QSOs), where the radiative power is by definition very high, objects with high radio luminosities form $\sim$10\% of the population, although it is not clear whether this is a stable phase or whether the radio-luminous phase is intermittent \citep[e.g.][]{2005ref57,2010intr152}. The remaining $\sim$90\% of quasars also produce radio emission \citep[][]{2013intr148} but this is not as strong as we observe in radio galaxies and the more radio-luminous quasars \citep[e.g.][]{1992lf5} and in some cases may be entirely due to star formation in the host galaxy \citep[e.g.][]{kimball+11,condon+13}.

 Traditionally in the literature quasars have been classified using radio and optical measurements as radio-loud (RL) or radio-quiet (RQ) quasars \citep[e.g.][]{1964ref176,Stocke+92}. However, these diagnostics are based on ratios of the radio luminosity to the optical luminosity, where the optical emission is a combination of emission from the accretion, optical jet and stars, and the radio emission may also be contaminated by the host galaxy and bears a complex relationship with the underlying jet power, depending on environment, time, and Doppler boosting among other factors. Such categorisations cannot therefore be expected to provide unambiguous information in all cases. Ideally we would classify these sources using their accretion and jet powers, but these, particularly the jet powers, are difficult to obtain observationally.

A great deal of attention has been devoted to the origin and physical reality of the RL/RQ difference in quasars with the goal of understanding whether there is a fundamental physical difference between quasars with and without strong radio emission. Such studies are complicated by issues of the definition of radio loudness and suffer from a number of problems. Firstly, as mentioned above, the classification ratios defined to date are not clear: a source can be classified as a RL quasar according to one classification and a RQ quasar in another\footnote{The standard definition was used to separate RL and RQ quasars in the literature is the ratio of 1.4-GHz radio luminosity to optical i-band luminosity: L$_\mathrm{{1.4-GHz}}$/L$_\mathrm{{i-band}}>1.$}. Secondly, the definition of radio loudness involves using fluxes (or luminosities) at whatever optical and radio bands are available, and the use of different bands may not give consistent results \citep[e.g.][]{Kellermann+89,Falcke+96,Stocke+92,Ivezic+02,Jiang+07}. Thirdly, the definition of radio loudness is often constructed for a particular sample that has a unique redshift or optical luminosity distribution. It is thus not surprising that studies that have focussed on the dichotomy in quasars in terms of radio loudness had contradictory conclusions. Some studies have suggested that there is a bimodality in the radio loudness of quasars \citep[e.g.][]{Ivezic+02,White+07} and others have disputed the reality of this bimodality \citep[e.g.][]{Falcke+96,White+00,Lacy+01,Brotherton+01,Cirasuolo+03,Cirasuolo+032,Balokovic+12}. Certainly it is the case that a good fraction of the quasars classified as RL and RQ in the literature present similar properties \citep[e.g.][]{Zamfir+8,Sulentic+2000}.

It has been generally thought that the RQ/RL difference involves the presence or absence of a relativistic jet. However, it should be noted that there may not be only one mechanism powering the radio emission and there are probably a number of sources in which the radio continuum might well be a combination of radio emission from small-scale jets as well as star formation \citep{Cirasuolo+03}. Sub-arcsec resolution is required to separate these two components: AGN and star formation. Plausible jet-generation mechanisms involve a rotating black hole and the accretion of magnetic flux \citep[e.g.][]{2009intr144}; for a review see \cite{2007intr145}. The dependency of radio loudness on different black hole and/or galaxy properties has therefore also been investigated. These properties include black hole mass, Eddington ratio, black hole spin, magnetic flux \citep{Sikora+Bagelman13}, galaxy morphology, and galaxy environment. \citep[][and references therein]{Sikora+07}. No firm observational conclusion has yet been reached. The situation is complicated by the fact that even objects with no discernible jet are expected to produce radio emission since star formation generates a galaxy-wide population of synchrotron-emitting cosmic rays and quasar hosts are expected to often be star-forming galaxies \citep[e.g.][]{2014ref138}. Some studies have suggested that the radio emission of RQ quasars comes from star formation \citep[e.g.][]{kimball+11,condon+13} while others argue that radio emission in these sources is due to AGN \citep[e.g.][]{Zakamska+16,SWhite+15,symeon16,White+17}. Again, these differences between different studies might be a consquence of the data used and the sample selection.

Another important aspect of these studies is the observed radio frequency. Because quasars are rare in the local Universe, studies of large samples of quasars across cosmic time require wide-area sky surveys. To date, studies of large samples of quasars have almost exclusively used radio surveys carried out at high ($>1$ GHz) radio frequencies such as Faint Images of the Radio Sky at Twenty Centimetres \citep[FIRST;][]{Becker+95} and the NRAO VLA Sky Survey \citep[NVSS;][]{Condon+98}. At high frequencies Doppler boosting might have an important effect on the measured radio luminosity for jetted sources, whereas low-frequency radio measurements are dominated by extended structures (lobes, plumes etc.) that are not Doppler-boosted. With new low-frequency radio interferometer arrays such as the Low Frequency Array \citep[LOFAR;][]{vanHaarlem2013}, the Murchison Widefield Array \citep[MWA;][]{Bowman+13,Tingay+13} and the Giant Metrewave Radio Telescope \citep[GMRT;][]{gmrt91} we are able to move towards lower radio frequencies, at which the effects of Doppler boosting can be minimised. A number of LOFAR surveys have been carried out over specific fields such as the Lockman hole field \citep{Mahony16}, the Bo\"otes field \citep{2016williams}, and the Herschel-Astrophysical Terahertz Large Area Survey/North Galactic Pole field \citep[H-ATLAS/NGP;][]{Hardcastle+16}. Recently, LOFAR has started observing the northern sky as part of the LOFAR Two-metre Sky Survey \citep[LoTSS;][]{Shimwell+17}, which provides unprecedented sensitivity ($\sim$ 70$\mu$Jy/beam) with a resolution of 6 arcsec; for optically thin synchrotron emission LoTSS is ten times deeper than the FIRST survey (assuming $\alpha=0.7$) and is also sensitive to extended emission that is invisible to FIRST. The $\sim 424$-deg$^2$ Hobby-Eberly Telescope Dark Energy Experiment (HETDEX)  \citep{Hill+8} Spring field has been chosen as the demonstrator field for LoTSS and is the widest area contiguous field available at this combination of sensitivity and frequency (120-168 MHz).

In this paper we use the LoTSS data release 1 (DR1) data over the HETDEX spring field and the LOFAR H-ATLAS/NGP survey to investigate the low-frequency radio properties of optically selected quasars from the Sloan Digital Sky Survey -- Baryon Oscillation Spectroscopic Survey \citep[SDSS-BOSS;][]{Paris+17}. In particular we concentrate on the radio loudness of quasars and its dependence on other galaxy and black hole parameters such as black hole mass, optical bolometric luminosity, radio luminosity, redshift, and Eddington ratio. Combining SDSS data with highly sensitive and high-resolution LOFAR observations in these fields we gather the largest sample of optically selected quasars detected at 144 MHz to date. The key value of our survey is that the LoTSS data are deep enough to allow direct detection of a significant percentage (approaching 50\%) of SDSS quasars at all redshifts. We show that the optically selected quasars present a wide range of radio continuum properties and their loudness does not appear to depend on the quasar nuclear properties. The results derived from this work highlight the impact of the deep and high-resolution low-frequency radio surveys, which foreshadow the compelling science cases for the Square Kilometre Array (SKA).

The layout of this paper is as follows. A description of the sample and data used in this work are given in Section \ref{sec:data}. The key results are given in Section \ref{sec:results}, in which we present the classification of quasars, discuss the dependency of the radio loudness parameter on various black hole or source parameters, and evaluate the far-infrared (far-IR)--radio correlation of quasars. In Section \ref{sec:discuss} we discuss our findings and compare with the literature. Finally, Section \ref{sec:discuss2} presents a summary of the results and conclusions drawn.

Throughout the paper we use the most recent Planck cosmology \citep{Planck+16}: $H_0=67.8$ km s$^{-1}$ Mpc$^{-1}$, $\Omega_{m}=0.308$ and $\Omega_{\Lambda}=0.692$. The radio spectral index $\alpha$ is defined in the sense $S \propto \nu^{-\alpha}$.

\section{Data}

\label{sec:data}

\subsection{The sample}

\label{sec:sample}

Our quasar sample is drawn from the SDSS quasar catalogue 14th data release \citep[DR14Q;][]{myers+15}, which includes all SDSS-IV/the extended Baryon Oscillation Spectroscopic Survey \citep[eBOSS;][]{blanton17} objects that were spectroscopically targeted as quasar candidates and that are confirmed as quasars via a new automated procedure combined with a partial visual inspection of spectra. The SDSS quasar target selection and quasar catalogue description are given in detail by \cite{Ross+12} and \cite{Paris18}. Quasars were targeted for spectroscopy by SDSS \citep{Richards+02} by selecting point sources that occupy a certain region in colour--colour space (far from the locus of stars) in optical colour--colour space (i.e optically selected or colour-selected). Additionally, point sources with radio emission from FIRST \citep{Becker+95} were targeted.

We started with a sample of 49,972 quasars over the HETDEX and H-ATLAS/NGP (over which we have far-IR data available) regions. Visual inspections of some sources were also performed (see Section \ref{sec:lofar-flux}). This process allowed us to identify quasars with false optical counterparts due to very close neighbouring sources. There are 47 (out of 49,972 objects) sources identified this way that were excluded from the sample. This left us with 49,925 quasars in the sample. The quasar catalogue includes sources selected purely based on matching quasar candidates to the FIRST catalogue within 2" (i.e radio selected and outside the colour selected space, 185 objects out of 49,925). In order not to be biased by the sample selection and with our interpretation of the results obtained in this work, we separated quasars selected by their optical colours from those selected using the FIRST survey match criterion, and evaluated these separately. It is also possible that 2-arcsec positional matching might miss extended sources without detected cores in the FIRST survey.

Whenever available, we used black hole masses estimated using the \ion{}{MgII} and \ion{}{CIV} emission line widths, optical bolometric luminosities (derived using the quasar luminosities at 1350, 3000, 5100 \AA) and Eddington ratios published by \cite{Shen11}. Otherwise, we used estimates of the same quasar properties from \cite{Kozlowski17}. In total out of 49,925 objects there are 30,897 quasars with nuclear properties measured. These values were used in some part of the analyses presented in this work. Table \ref{det-table} provides some sample properties and detection rates.

\begin{figure}
\begin{center}
\scalebox{0.8}{
\begin{tabular}{c}
\hspace{-1em}\includegraphics[width=11.5cm,height=11.5cm,angle=0,keepaspectratio]{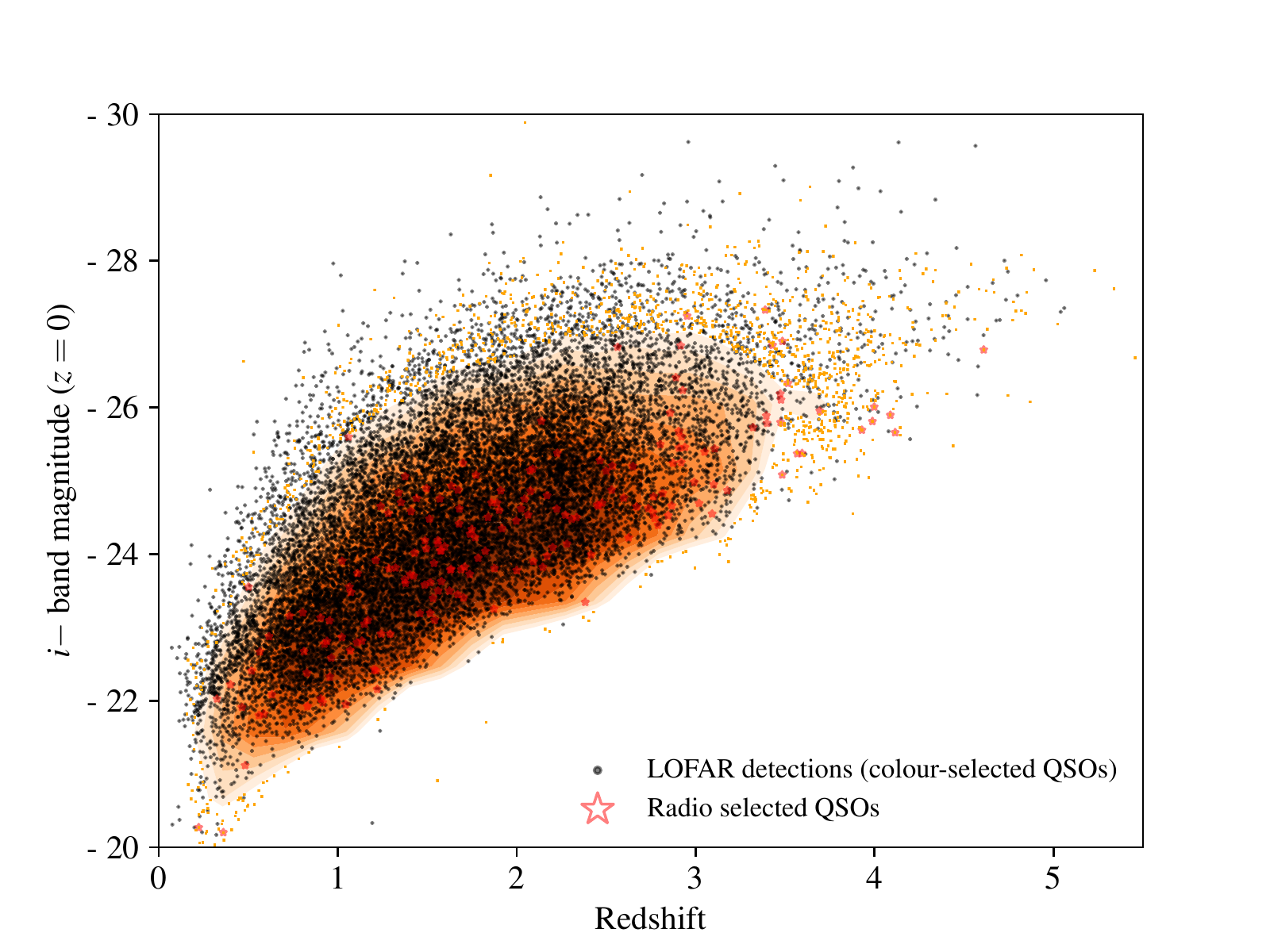}\\
\end{tabular}}
\caption{Distribution of $i$-band absolute magnitudes of quasars
  over the HETDEX and H-ATLAS/NGP fields, corrected for extinction and
  scaled to $z=0$, as a function of redshift. Black points (the 3$\sigma$ detections) and orange contours (LOFAR limits) show optically selected quasars and open red stars are FIRST selected quasars. \label{mag-plt}}
\end{center}
\end{figure}

\begin{table*}
\caption{Detection statistics and sample properties. \label{det-table}}
\begin{center}
\begin{tabular}{lr}
\hline
Sample over the HETDEX $\&$ H-ATLAS/NGP fields & Number\\
\hline
All quasar sample over the HETDEX and H-ATLAS/NGP&49,972\\
Quasars with false optical counterparts&47\\
Quasars with correct optical counterparts&49,925\\
Quasars selected by their optical colours, regardless of their FIRST counterparts (CS)& 49,740\\
Quasars selected by the FIRST match criterion (FS) &185\\
Quasars selected by their optical colours, regardless of their FIRST counterparts, detected by LOFAR (CSD)&16,077\\
Quasars selected by their optical colours, regardless of their FIRST counterparts, LOFAR limits (CSL)&33,663\\
Quasars selected by the FIRST match criterion detected by LOFAR (FSD) &178\\
Quasars selected by their FIRST match criterion, LOFAR limits (FSL)&7\\
Quasars that have measured nuclear properties&30,897\\
\hline
  \end{tabular}
  \end{center}
\end{table*}

\subsection{Optical data} 

\label{sec:optdata}

The DR14Q quasar catalogue contains $i$-band absolute magnitudes ($k$-corrected to $z=2$), SDSS magnitudes and fluxes at $u$,$g$,$r$,$i$,$z$ bands, together with various properties derived from these. These measurements were used in the analyses presented in this work. Additionally we obtained extinction-corrected $i$-band absolute magnitudes $k$-corrected to $z=0$ by converting the $i$-band absolute magnitudes $k$-corrected to $z=2$ using the following conversion given by \cite{Richards+6}:

\begin{equation}
\label{eq:opt_mag}
M_{i}(z=0) = M_{i}(z=2)+2.5(1+\alpha)\log(1+z)
,\end{equation}

where $M_{i}(z=0)$ is the absolute magnitude $k$-corrected to $z=0$, $M_{i}(z=2)$ is the absolute magnitude $k$-corrected to $z=2,$ and $\alpha$ is the optical spectral index. We used the canonical value of $\alpha=-0.5$ \citep{Richards+6}.

The derived absolute magnitudes were used to calculate $i$-band
luminosities using the following relation: $L_{i} = L_{\odot} \times
10^{-0.4(M_{i} - m_{\odot})}$, where $L_{i}$ is the $i$-band
luminosity, M$_{i}$ is the $i$-band absolute magnitude, and
$L_{\odot}$ and $m_{\odot}$ are the solar luminosity (3.8270$\times$10$^{26}$ W) and solar
magnitude in the optical band (which is 4.58), respectively. The $i$-band absolute magnitude (scaled to $z=0$) distribution of the quasars selected over the fields can be seen in Fig.\ref{mag-plt}.

\subsection{Radio data} 

\subsubsection{Flux densities at 144 MHz}

\label{sec:lofar-flux} 

As mentioned above we combined the LOFAR data over the HETDEX Spring and H-ATLAS/NGP fields. The HETDEX Spring field (right ascension 10h45m00s to 15h30m00s and declination $\ang{45}$00$^{\prime}$00$^{\prime\prime}$ to $\ang{57}$00$^{\prime}$00$^{\prime\prime}$) was observed with LOFAR as part of LoTSS. This field was targeted as it is a large contiguous area at high elevation for LOFAR, whilst having a large overlap with the SDSS \citep{york00} imaging and spectroscopic data. Importantly, this field also paves the way for using HETDEX data to provide emission-line redshifts for the LOFAR sources and prepares for the WEAVE-LOFAR\footnote{http://www.ing.iac.es/weave/weavelofar/} survey, which will measure spectra of more than 106 LOFAR-selected sources \citep{smith+16}. The region was also chosen because HETDEX is a unique survey that is very well matched to the key science questions that the LOFAR surveys project aims to address. In particular, the ability to obtain [O II] redshifts up to z$\sim$0.5 is well matched to the LOFAR goal of tracking the star formation rate density using radio continuum observations. The creation of the radio images is described by Shimwell et al. (2018). Radio flux densities at 144 MHz for all SDSS quasars in our sample were directly measured from the final full-bandwidth LOFAR maps (in total 49,925 objects). We performed Gaussian fitting procedure to the point sources to extract their fluxes using {\sc astropy/photutils} \citep{larry_bradley_2017_1039309}. The noise-based uncertainties on these flux densities were estimated using the LOFAR r.m.s. maps. 

In order to obtain the LOFAR fluxes of sources showing extended emission (sources showing clear extended emission divided into multiple components; see Appendix for some examples), we firstly selected quasars detected at 5$\sigma$ at 144 MHz by cross-matching the quasar catalogue to the catalogue produced by PyBDSF \citep{Mohan+Rafferty15} within a 5-arcsec match radius; this radius was chosen taking into account the resolution of LOFAR maps. These were inspected visually to select sources that present extended structures; 773 sources were selected in this way. We also checked whether we missed any source by our first matching process by cross-matching the quasar catalogue with the value-added catalogue, which was constructed by a process involving visual inspection of sources (Williams et al. 2018; Duncan et al. 2018). This showed that using our first matching process we did not miss any source. We then used the value-added LOFAR/HETDEX catalogue, which has total fluxes of sources with multiple components (Williams et al. 2018; Duncan et al. 2018).

The same process was applied to the LOFAR data over the H-ATLAS/NGP region using the best available LOFAR maps of the field \citep{Gurkan+18}. We currently do not have a value-added catalogue for the H-ATLAS/NGP field. Therefore, we selected sources detected at 5$\sigma$ at 144 MHz in the same way described above, then identified by visual inspection the sources showing extended emission. The total fluxes of sources with multiple components (83 quasars) were derived by combining individual component fluxes. In total we selected 856 sources with extended radio emission over the HETDEX and H-ATLAS/NGP fields.

\begin{figure}
\begin{center}
\scalebox{0.8}{
\begin{tabular}{c}
\hspace{-1em}\includegraphics[width=11.5cm,height=11.5cm,angle=0,keepaspectratio]{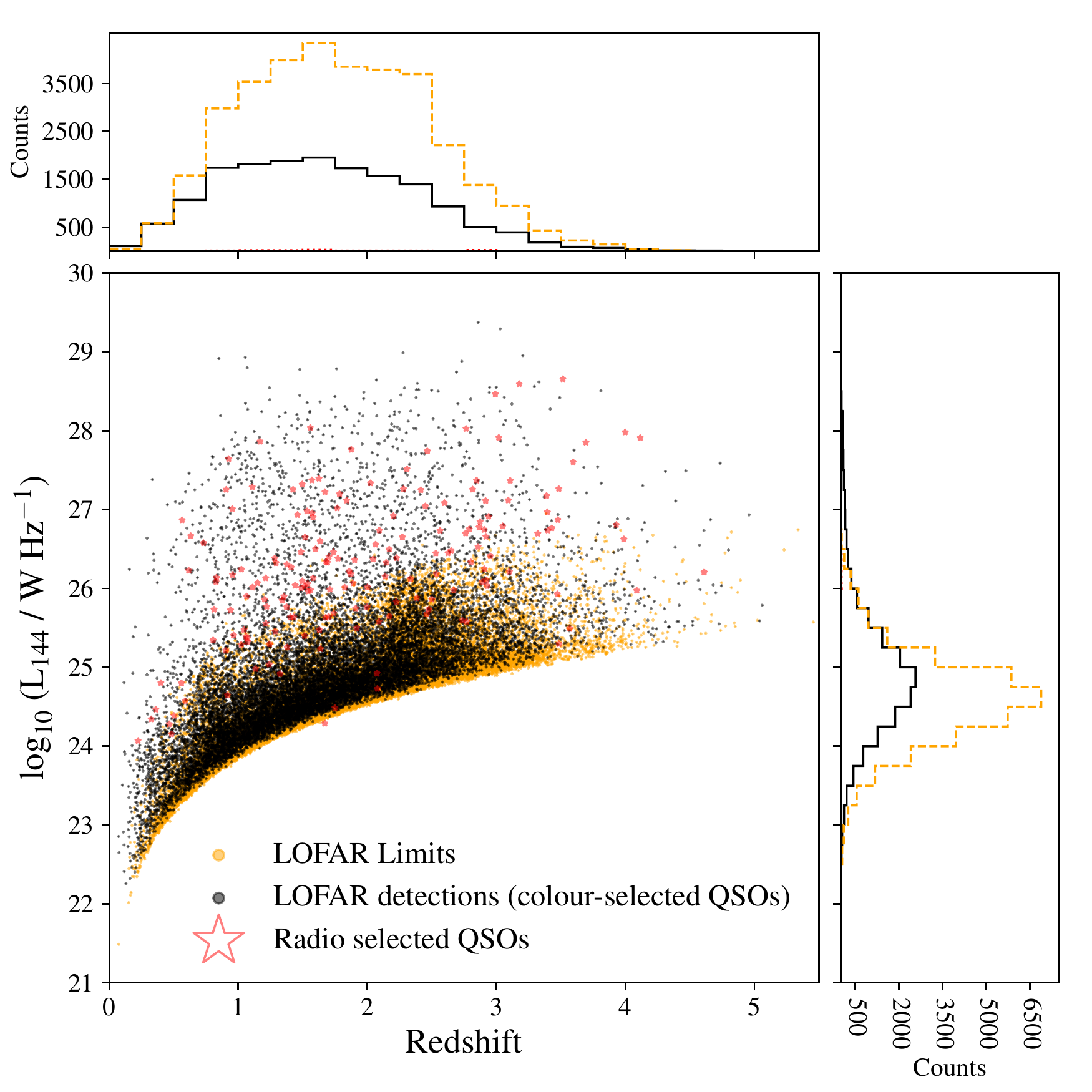}\\
\end{tabular}}
\caption{Distribution of 144-MHz luminosity of quasars as a function of their redshifts. Yellow dots indicate quasars that were not detected in LOFAR at 3$\sigma$, black dots represent the 3$\sigma$ detections, and red open stars radio-selected quasars. In the top and right panels the quasars detected in LOFAR are shown with solid lines and the non-detections with dashed lines.\label{lfr-plt}}
\end{center}
\end{figure}

To convert 144-MHz flux densities to $k-$corrected 144-MHz luminosities ($L_{144}$ in W Hz$^{-1}$) we adopt a spectral index $\alpha=0.7$ \cite[the typical value found by][]{Hardcastle+16}. The distribution of 144-MHz luminosity of quasars as a function of their redshifts is seen in Fig. \ref{lfr-plt}.

\subsubsection{Flux densities at 1.4 GHz} 

\label{sec:spix}

We obtained the FIRST \citep{Becker+95} images and r.m.s. maps of the HETDEX and H-ATLAS/NGP fields. As for the LOFAR flux density measurements, we measured the flux densities at the source positions 
by fitting a Gaussian model. Uncertainties on these flux densities were estimated in the same way as for the LOFAR flux errors using the 1.4-GHz r.m.s. maps. The $k-$corrected 1.4-GHz luminosities of the sources in the sample ($L_{1.4}$ in W Hz$^{-1}$) were estimated using these flux densities and a spectral index $\alpha = 0.7$ at the spectroscopic redshift.

\begin{figure}
\begin{center}
\scalebox{1.1}{
\begin{tabular}{c}
\hspace{-3em}\includegraphics[width=11.5cm,height=11.5cm,angle=0,keepaspectratio]{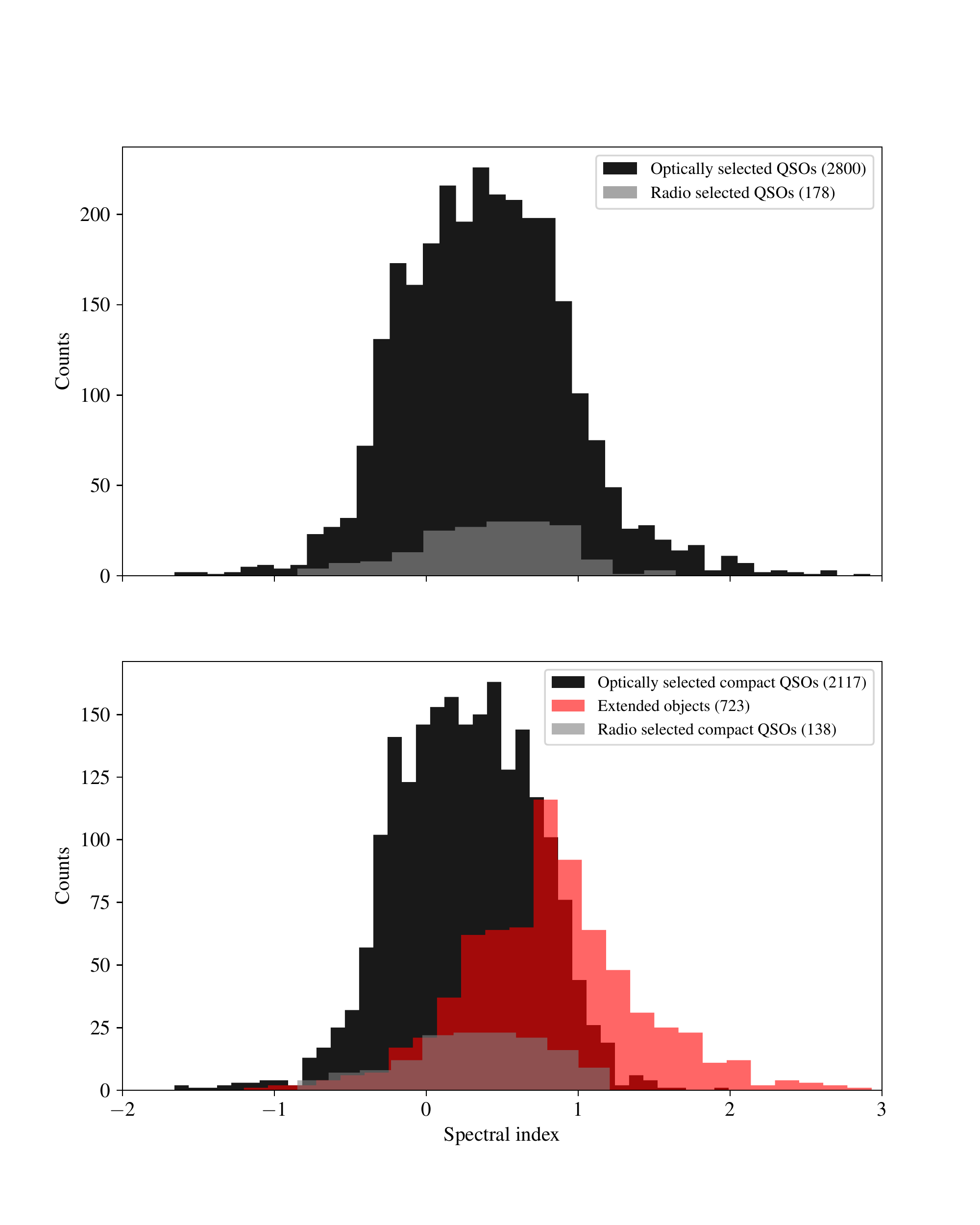}\\
\end{tabular}}
\caption{Top: Distribution of $\alpha$ of quasars calculated using LOFAR and FIRST flux densities. The distribution is shown for quasars detected by both LOFAR and FIRST, separately for optically selected and radio selected quasars. Bottom: The distribution of $\alpha$ of optically, radio selected and objects presenting extended structures. Optimal bin widths were selected using the Knuth rule \citep{deeming75,astroMLText}. \label{index-dist}}
\end{center}
\end{figure}

We do not use spectral indices estimated using LOFAR and FIRST flux densities as this is not a true estimate of the population spectral index, and the biases are complex because the LOFAR data are deeper than FIRST. Additionally FIRST is not as sensitive as LOFAR to extended emission so it might be missing some flux from extended souces. However, we simply checked the distribution of $\alpha$ of quasars detected by both telescopes. As LOFAR is deeper than FIRST we expect most sources, which are detected by FIRST, to be detected by LOFAR as well. In Fig. \ref{index-dist} we show the $\alpha$ distribution of quasars detected by both FIRST and LOFAR. We show these quasars using different colours in order to see differences
between quasars selected based on different criterion. Quasars that show extended emission, identified by visual inspection, were also shown separately. This figure reveals the difference between optically and radio selected quasars in terms of their spectral indices: optically selected quasars (excluding extended sources) have flatter spectral indices (mean = $0.26\pm0.01$, median = $0.26\pm0.02$) than radio selected quasars (excluding extended sources, mean = $0.33\pm0.04$, median = $0.36\pm0.06$). Extended sources, as expected, are the steepest radio spectra in comparison to point-like sources (mean = $0.82\pm0.02$, median = $0.81\pm0.02$) because FIRST was missing some of the extended emission. As mentioned above this is just a simple check and these values should be used with a caution. This is because a large percentage of sources are not detected by FIRST which have fainter flux densities at 1.4 GHz than at 150 MHz.


\subsection{Far-infrared data} 

\label{sec:firdata}

$\textit{Herschel}$-ATLAS  provides imaging data for the $\sim$142-deg$^2$ NGP field using the Photo-detector Array Camera and Spectrometer \cite[PACS at 100 and 160 $\mu$m; ][]{2010ref10,poglitsch10} and the Spectral and Photometric Imaging Receiver \citep[SPIRE at 250, 350, and 500 $\mu$m;][]{2010ref11,2011ref145,2016valiante}. To derive a maximum-likelihood estimate of the flux densities at the positions of objects in the SPIRE bands whether formally detected or not, the point spread function (PSF)-convolved H-ATLAS images were used for each source together with the errors on the fluxes. Further details of the flux measurement method are given by \cite{2010ref7,2013ref9}.

In order to estimate 250 $\mu$m luminosities ($L_{250}$ in W Hz$^{-1}$) for our sources we assumed a modified black-body spectrum for the far-IR spectral energy distribution (SED; using both SPIRE and PACS bands); we fixed the emissivity index $\beta$ to 1.8 [the best-fitting value derived by \cite{2013ref9} and \cite{2013ref87} for sources in the H-ATLAS] and obtained the best-fitting temperatures, integrated luminosities ($L_{\mathrm{IR}}$), and rest-frame luminosities at 250 $\mu$m ($L_{250}$) by minimising $\chi^2$ for all sources with significant detections. To calculate the 250 $\mu$m $k$-corrections the same emissivity index and the mean of the best-fitting temperatures were used for quasars. These corrections were included in the derivation of the 250 $\mu$m luminosities that are used in the remainder of the paper.


\subsection{Radio loudness of quasars}

\label{sec:radio_class}

There are various diagnostics using radio and optical measurements to classify quasars as RL or RQ. As noted in Section \ref{sec:intro}, traditional classifications are mainly based on the ratio of a radio measurement (flux density or luminosity) to an optical measurement \citep[e.g.][]{Kellermann+89,Falcke+96,Stocke+92,Ivezic+02,Jiang+07}. We define the radio loudness parameter for our quasar sample using the ratio of $L_{144}$ to $i$-band luminosity. We use $i$ band for our analysis for several reasons: (i) fluxes that are measured by redder passbands are less sensitive to the part of the galaxy spectrum that is affected by recent star formation, (ii) redder passbands suffer less dust extinction, and (iii) the $i$ band has been previously used with FIRST flux densities to estimate the radio loudness for quasars \citep[e.g.][]{Ivezic+02,2014ref143}. Therefore, $i$-band magnitudes have been taken as the quantitative estimate for the optical luminosity (a good tracer of the accretion luminosity) and absolute magnitude. The radio loudness parameter ($\rr$ hereafter) is then defined as follows:

\begin{equation}
\rr = \log_{10}\left(\frac{L_{radio}}{L_{optical}}\right) = \log_{10}\left(\frac{L_{\mathrm{144\,MHz}} / \mathrm{W\ Hz^{-1}}} {L_{\mathrm{i\,band}} / \mathrm{W\ Hz^{-1}}}\right)
.\end{equation}

\begin{figure*}
\begin{center}
\scalebox{2}{
\begin{tabular}{c}
\hspace{-3em}\includegraphics[width=11.5cm,height=11.5cm,angle=0,keepaspectratio]{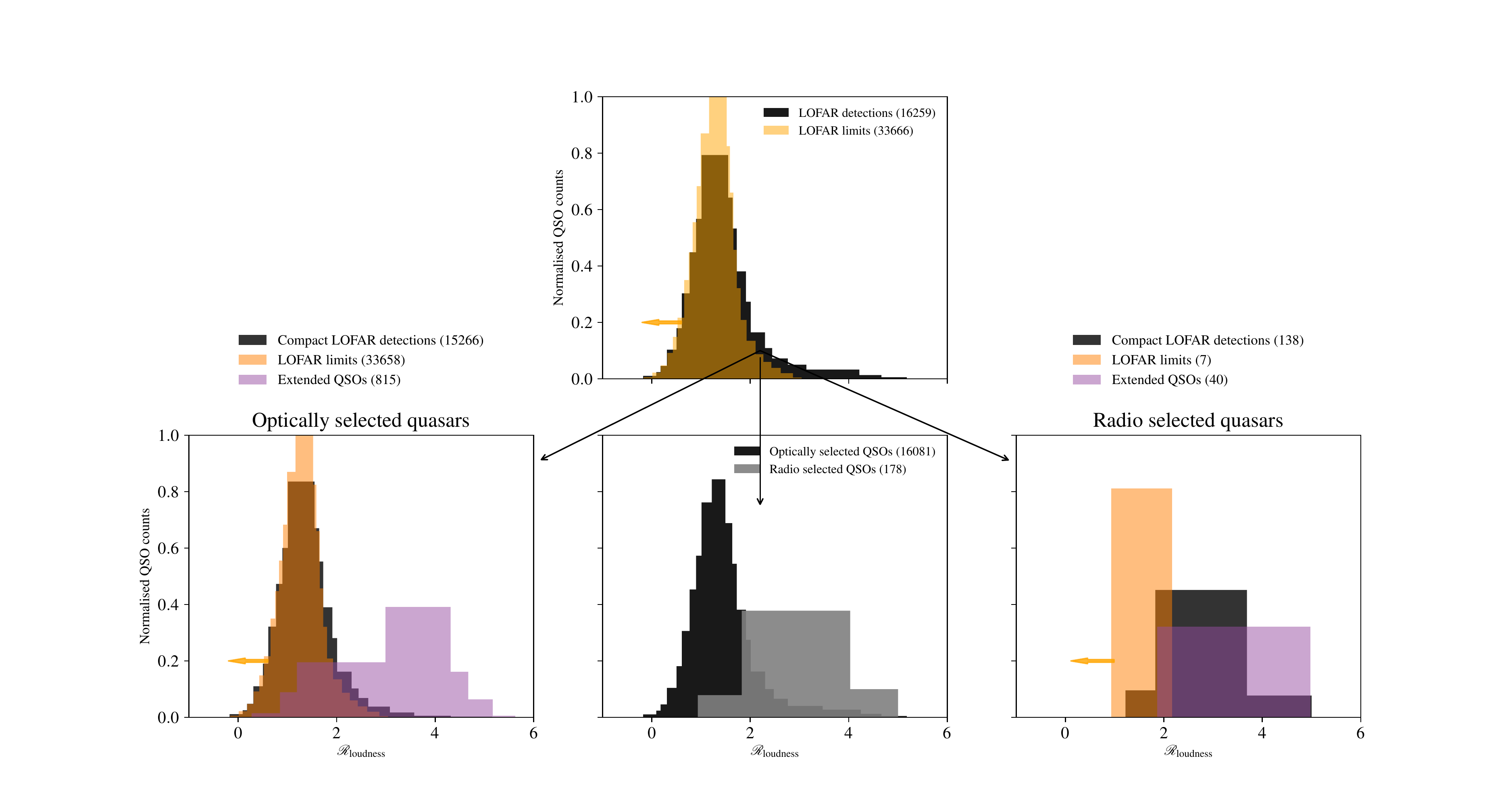}\\
\end{tabular}}
\caption{Top: Histogram of radio loudness parameter $\rr$ derived using $L_{144}$
  and SDSS $i$-band measurements of the whole sample. Black: LOFAR
  3$\sigma$ detections; orange: LOFAR limits. Bottom middle: Histogram of $\rr$ of the LOFAR detected sources, split based on their selections (i.e. optically selected and radio selected). Grey: Radio selected quasars; black: optically selected quasars. Bottom left: Histogram of $\rr$ of optically selected quasars. Black: LOFAR 3$\sigma$ detections; orange: LOFAR limits; and purple: extended sources. Bottom right: Histogram of $\rr$ quasars selected based on their match to the FIRST counterparts. Colours as for the bottom left panel. Optimal bin widths were selected using the Bayesian blocks formalism given by \cite{astroMLText}. \label{loudness}}
\end{center}
\end{figure*}

In the top panel of Fig. \ref{loudness} we show the radio loudness parameter $\rr$ histogram of the full sample, split by their detection properties. In the bottom middle panel of Fig. \ref{loudness} we show a histogram of $\rr$ for optically and radio selected sources, detected in LOFAR. We further split the sample by their selection criterion (i.e. optically or radio selected). In the bottom left panel we indicate optically selected quasars detected by LOFAR, limits, and objects that show indication of extended emission, identified by visual inspection. In the bottom right panel we show the $\rr$ histogram of quasars selected by their match to the FIRST counterparts detected in LOFAR, limits and those that show indication of extended emission. The median $\rr$ and 144-MHz flux densities with their bootstrap errors are given in Table \ref{det-table2}.

\begin{table*}
\caption{Detection statistics, the 144-MHz flux density and $\rr$ properties of quasars over the HETDEX and H-ATLAS/NGP fields. \label{det-table2}}
\centering
\begin{tabular}{ccccc}
\hline 
&\multicolumn{2}{c}{Sample over the HETDEX field} & \multicolumn{2}{c}{Sample over the H-ATLAS/NGP field} \\
\hline 
&Detections & Limits  & Detections & Limits \\
\hline 
Number&13982&28763&2272&4908\\
Median 144-MHz flux density (mJy)&$0.42 \pm 0.01$ & $0.21 \pm 0.01$&$1.67 \pm 0.06$&$0.93 \pm 0.02$\\
Median $\rr$ &$1.31 \pm 0.01$&$1.23 \pm 0.01$&$1.82 \pm 0.02$ &$1.74 \pm 0.01$\\
\hline
\end{tabular}
\end{table*}

\section{Results}

\label{sec:results}

\subsection{Evaluation of the radio loudness in SDSS quasars}

As can be seen in the top middle panel of Fig. \ref{loudness}, the quasars detected by LOFAR and limits span a similar range of $\rr$, though detections have a tail of high $\rr$. As mentioned in Section \ref{sec:sample} in order not to be biased by the selection method of quasars we separate the sample based on their selections and evaluate their $\rr$ distribution separately. The bottom middle panel shows the $\rr$ histogram of optically and radio-selected quasars detected by LOFAR. Optically selected and LOFAR-detected quasars have $\rr$ values around 1.6, whereas this is around 3.0 for radio-selected quasars. Evaluation of the limits shows that optically selected quasar limits have similar $\rr$ values to the detections  (bottom left panel in Fig. \ref{loudness}). Sources that show extended structures have much higher $\rr$, typically around 3.5. There are not many quasars selected by the FIRST survey match criterion that are not detected by LOFAR. Most of these sources are found to be at the edge of LOFAR pointings where the noise is higher than the beam centre. There are only seven sources that meet this condition. Radio-selected quasar limits also have similar $\rr$ values to optically detected quasar limits. Similarly quasars with extended structures present the highest loudness estimates, which peak around 4.0. 

The shape of radio loudness histograms of this kind has been used in earlier studies to understand whether there are two distinct quasar populations \citep[e.g.][]{Cirasuolo+03,Ballo+12,Balokovic+12}. Some authors \citep[e.g.][]{White+07} have claimed to see a bimodal distribution of this parameter, which would be taken as evidence for two different radio emission mechanisms in these sources;  radiation in RL objects is due to jet activity and in the RQ objects due to coronal activity, winds, or star formation. Firstly, in the top panel of Fig. \ref{loudness}, however, there is no significant evidence for a bimodal distribution when we consider the full sample of quasars and this conclusion is not affected by the large number of upper limits on $\rr$. In the bottom middle panel, however, the distribution has two peaks and there is some overlap as we split the sample based on their selections. 

To quantitatively evaluate the $\rr$ distribution of quasars we fitted the data using two different models: a single Gaussian and a Gaussian mixture model with two components using Markov chain Monte Carlo (MCMC) sampling, a routine provided by \cite{astroMLText}. We then computed the odds ratio for the models. This analysis was implemented using the following samples: 

\begin{itemize}
\item Quasars detected by LOFAR, including both optically and radio-selected objects. The odds ratio $=O_{21}=1.01.$
\item Quasars selected by their optical colours and detected by LOFAR (naturally this sample includes all extended objects). The odds ratio $=O_{21}=1.12.$
\item Quasars selected by the FIRST match criterion and detected by LOFAR. The odds ratio $=O_{21}=5.67.$ 
\end{itemize}

Computed odds ratios for the two aforementioned samples are very close to the unity and therefore are inconclusive: neither model is favoured by the data. However, this is 5.67 for radio-selected quasars, albeit not strong, this result suggests that a mixture Gaussian model is weakly favoured by the data over a single Gaussian model, although the number of sources in this sample is just 178.

\begin{figure*}
\begin{center}
\scalebox{1.6}{
\begin{tabular}{c}
\hspace{-1em}\includegraphics[width=11.5cm,height=11.5cm,angle=0,keepaspectratio]{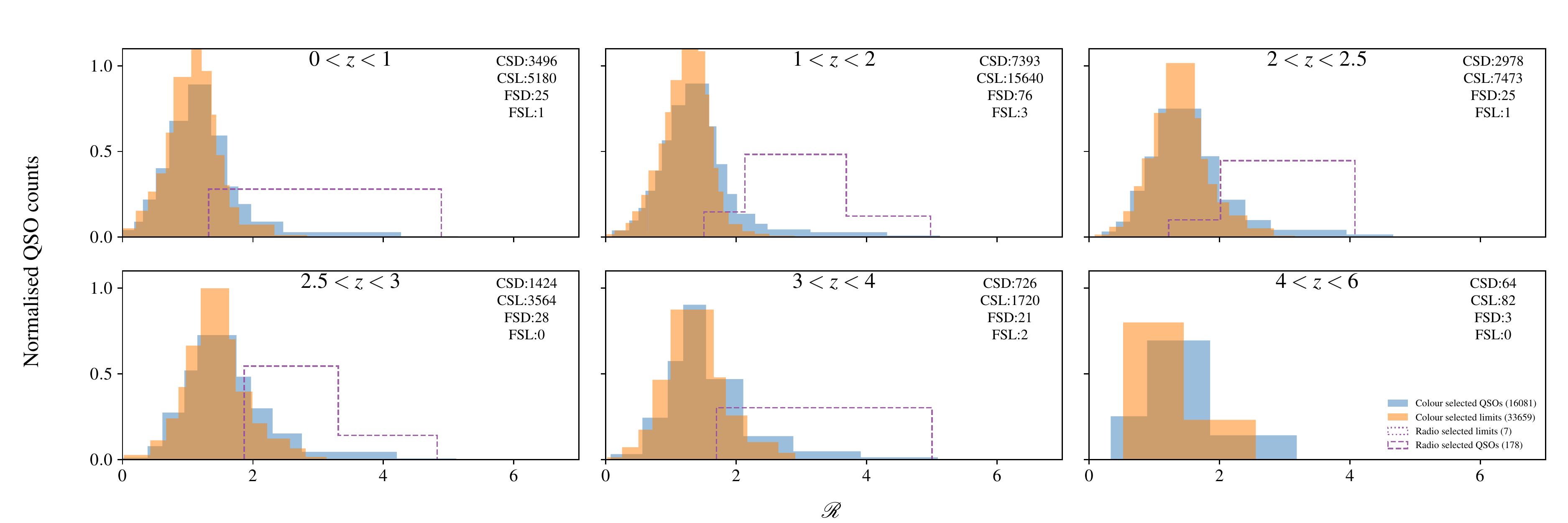}\\
\end{tabular}}
\caption{Histogram of radio loudness parameter for various redshift bins. CSD: quasars selected by their optical colours and detected in LOFAR, CSL: quasars selected by their optical colours do not have $3 \sigma$ detections in LOFAR, FSD: radio selected quasars detected in LOFAR, FSL: radio selected quasars which are not detected at $3 \sigma$ in LOFAR. Optimal bin widths were selected using the Bayesian blocks formalism described by \cite{astroMLText}. \label{loudness-z}}
\end{center}
\end{figure*}

In Fig. \ref{loudness-z} we show the $\rr$ distribution of quasars in different $z$ bins to evaluate its evolution across cosmic time. The overall picture is very similar to Fig. \ref{loudness}. Even though both measured radio and optical luminosities depend strongly on redshift, the bulk of the quasar population has similar $\rr$ values, ranging between 1.3 and 2.0, and radio-selected sources show higher $\rr$ estimates (ranging from $\rr \sim 1$ to $\rr \sim 3.5$). In Table \ref{t-average-R} we show the mean and median $\rr$ estimates of optically and radio-selected quasars for different redshift bins. 


\begin{table*}
\caption{Mean and median $\rr$ estimates of quasar detections and limits for different $z$ bins and their bootstrap errors. \label{t-average-R}}
\begin{center}
\begin{tabular}{lrrrrr}
\hline
Category & $z$ bins &Mean $z$ &N& Median $\rr$ & Mean $\rr$\\ 
\hline
LOFAR detections (colour-selected) & $0<z<1$ & $0.70$ & $3496$ &$1.17 \pm 0.01$& $1.27 \pm 0.014$\\
& $1<z<2$ & $1.50$ & $7393$ &$1.36 \pm 0.01$& $1.47 \pm 0.009$\\
& $2<z<2.5$ & $2.24$ & $2978$ &$1.47 \pm 0.01$& $1.59 \pm 0.014$\\
& $2.5<z<3$ & $2.70$ & $1424$ &$1.53 \pm 0.02$& $1.65 \pm 0.020$\\
& $3<z<4$ & $3.30$ & $726$ &$1.47 \pm 0.02$& $1.63 \pm 0.029$\\
& $4<z<6$ & $4.33$ & $64$ &$1.32 \pm 0.09$& $1.46 \pm 0.087$\\
& $0<z<6$ & $1.66$ & $16081$ &$1.35 \pm 0.01$& $1.47 \pm 0.006$\\
LOFAR limits (colour-selected)& $0<z<1$ & $0.75$ & $5180$ &$1.09 \pm 0.01$& $1.09 \pm 0.006$\\
& $1<z<2$ & $1.51$ & $15640$ &$1.26 \pm 0.00$& $1.23 \pm 0.003$\\
& $2<z<2.5$ & $2.25$ & $7473$ &$1.39 \pm 0.00$& $1.40 \pm 0.005$\\
& $2.5<z<3$ & $2.71$ & $3564$ &$1.42 \pm 0.01$& $1.44 \pm 0.008$\\
& $3<z<4$ & $3.30$ & $1720$ &$1.33 \pm 0.01$& $1.37 \pm 0.013$\\
& $4<z<6$ & $4.44$ & $82$ &$1.14 \pm 0.05$& $1.26 \pm 0.056$\\
& $0<z<6$ & $1.78$ & $33659$ &$1.28 \pm 0.00$& $1.27 \pm 0.003$\\
LOFAR detections (radio-selected)& $0<z<1$ & $0.69$ & $25$ &$2.74 \pm 0.51$& $2.99 \pm 0.208$\\
& $1<z<2$ & $1.49$ & $76$ &$2.87 \pm 0.13$& $2.97 \pm 0.096$\\
& $2<z<2.5$ & $2.24$ & $25$ &$2.90 \pm 0.24$& $2.93 \pm 0.156$\\
& $2.5<z<3$ & $2.80$ & $28$ &$2.79 \pm 0.24$& $2.87 \pm 0.151$\\
& $3<z<4$ & $3.44$ & $21$ &$2.97 \pm 0.38$& $3.20 \pm 0.204$\\
& $4<z<6$ & $4.27$ & $3$ &$1.97 \pm 1.08$& $2.61 \pm 0.719$\\
& $0<z<6$ & $1.96$ & $178$ &$2.86 \pm 0.07$& $2.97 \pm 0.065$\\
LOFAR limits (radio-selected)& $1<z<2$ & $1.66$ & $3$ &$1.53 \pm 0.18$& $1.63 \pm 0.117$\\
& $0<z<6$ & $2.08$ & $7$ &$1.53 \pm 0.14$& $1.56 \pm 0.156$\\
\hline
\end{tabular}
\end{center}
\end{table*}

\subsection{Relation between $\rr$ and $z$, accretion and black hole mass}

\label{sec:z-dist}

\begin{figure}
\begin{center}
\scalebox{0.9}{
\begin{tabular}{c}
\hspace{-1em}\includegraphics[width=11.5cm,height=11.5cm,angle=0,keepaspectratio]{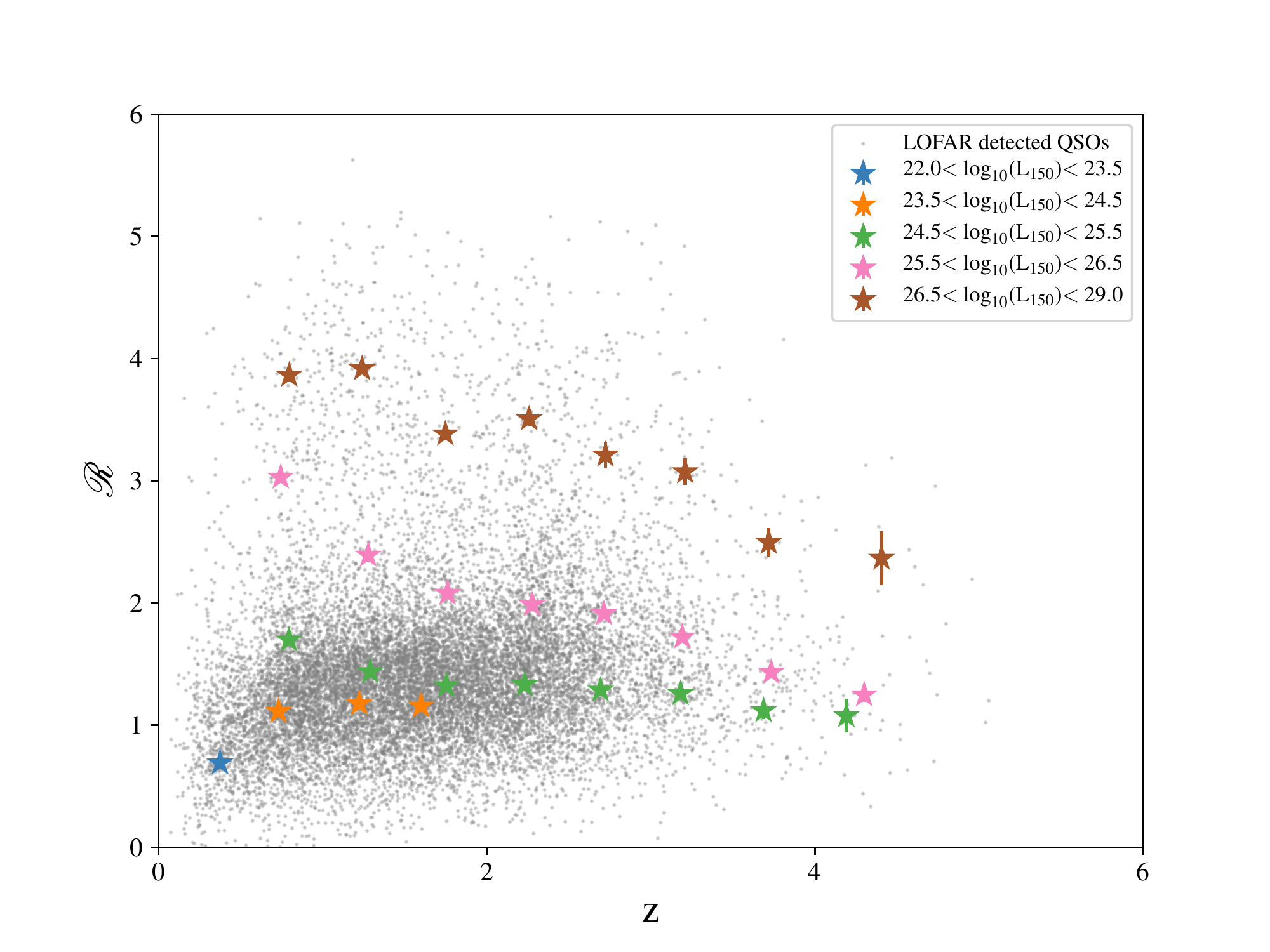}\\
\end{tabular}} 
\caption{Dstribution of $\rr$ as a function of redshift for quasars detected by LOFAR, indicated as grey points. Median-$\rr$ stacks are also shown for $z$ and $L_{144}$ bins with their bootstrap errors. \label{r-z-lofar-single}}
\end{center}
\end{figure}

 We now examine the relation between loudness parameter $\rr$ and redshift, black hole mass and accretion in our quasar sample. In Fig. \ref{r-z-lofar-single} we show the distribution of $\rr$ as a function of redshift for only the optically selected quasars detected by LOFAR. We also estimated median stacks for $\rr$ for various $z$ and $L_{144}$ bins. Evaluation of this figure indicates that although there is a spread in $\rr$ across cosmic time for quasars, the typical value of $\rr$ for low radio luminosity quasars remains more or less constant for a wide $z$ range. However, including radio luminosity information shows that there is a slight decrease in $\rr$ as a function of redshift for quasars with high radio luminosities [$25.5<\log_{10}$ ($L_{\mathrm{144}}$)$ <29.0$]. The decrease in $\rr$ with $z$ for high-power sources is consistent with expectations if these are powered by jets: owing to to increases in the inverse-Compton losses with increasing redshift, for a given jet power the radio luminosity of high power objects that we observe are lower at high redshifts than at low redshifts \citep[e.g.][]{Hardcastle18}. 

\begin{figure*}
\begin{center}
\scalebox{1.9}{
\begin{tabular}{c}
\hspace{-4em}\includegraphics[width=11.5cm,height=11.5cm,angle=0,keepaspectratio]{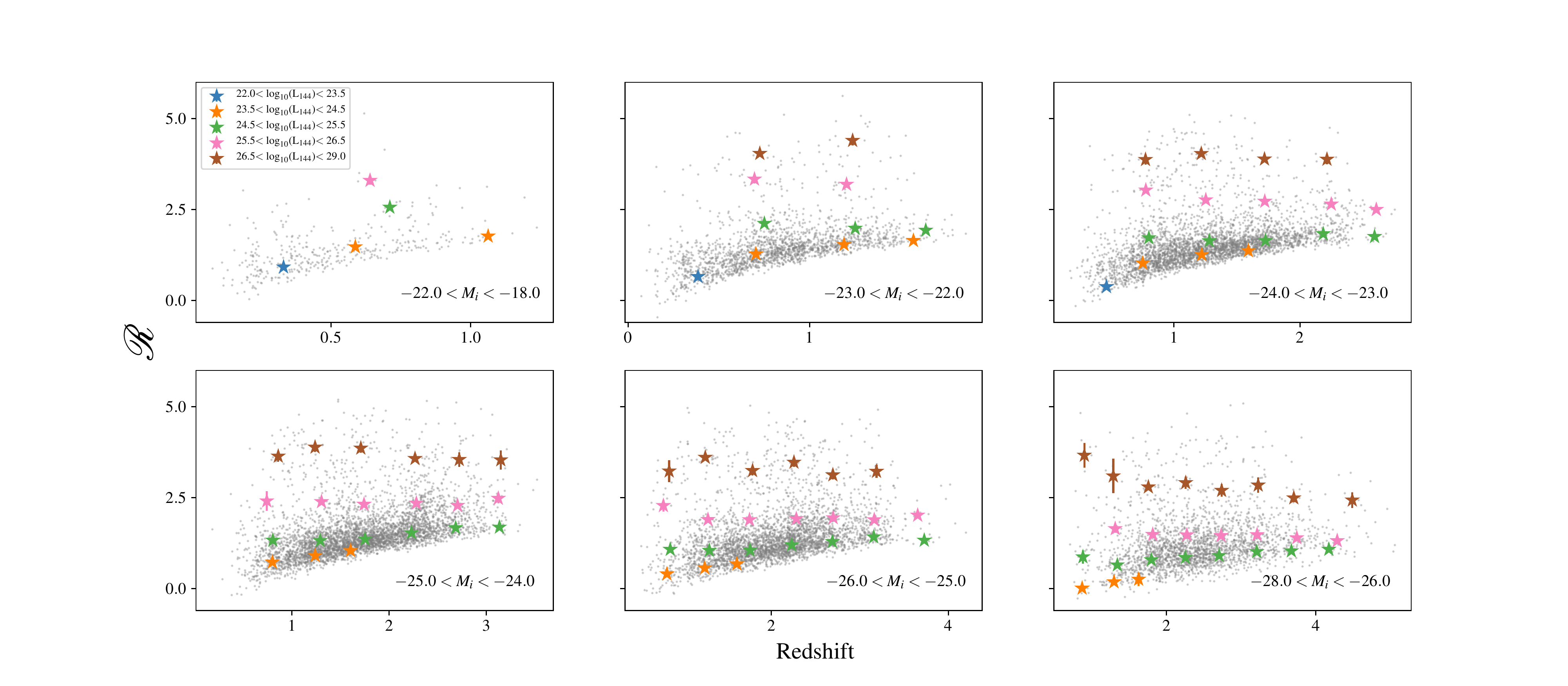}\\
\end{tabular}} 
\caption{Distribution of $\rr$ as a function of redshift in various absolute magnitude bins for quasars detected by LOFAR, indicated as grey points. Median-$\rr$ stacks are also shown for $z$ and $L_{144}$ bins with their bootstrap errors. \label{r-z-lofar}}
\end{center}
\end{figure*}

We also divide the sample in six absolute $i$-band magnitude bins and evaluate the relation between $\rr$ and $z$. Results of this analysis are shown in Fig. \ref{r-z-lofar}. Since we constrain the sample to LOFAR detected objects in each bin we do not have many sources to reveal the actual relation between $z$ and $\rr$ for the first two magnitude bins. In these panels we see that $\rr$ goes down with increasing redshift and correspondingly increasing $i$-band magnitude. This trend might be driven by the dependency of $M_{i}$ on z. We test this by performing partial correlation analysis between $\rr$ and $L_{\mathrm{i\,band}}$ for controlling the effect of $z$ for quasars detected in LOFAR (see Table \ref{spearman}). We find relatively strong anti-correlation between $\rr$ and $L_{\mathrm{i\,band}}$ whilst taking away the effect of redshift. Results of the partial correlation for LOFAR detected quasars in each $i$-band magnitude bin showed that the strength of this anti-correlation is increasing (from -0.16 to -0.23 with $p<0.0001$) with increasing magnitude (correspondingly increasing redshift). It is also interesting to investigate if there is any genuine correlation between $\rr$ and redshift when we control for the effect of $i$-band absolute magnitude. Even taking away the effect of $M_{i}$ we find a positive correlation between $\rr$ and $z$ (0.5, $p<0.0001$) for quasars detected by LOFAR and selected by their colours. All these results suggest that what we see might be a selection effect; i.e. we are not able to sample optically bright quasars. 

As mentioned in Section \ref{sec:intro} a number of different black hole parameters have been used to explain why some galaxies are bright in radio and show extended emission while most of them are RQ (but not radio silent). Two of these are the black hole mass and the Eddington ratio. In this section we explore the dependence of the radio loudness on these parameters.

\begin{figure*}
\begin{center}
\scalebox{1.7}{
\begin{tabular}{c}
\hspace{-1em}\includegraphics[width=11cm,height=11cm,angle=0,keepaspectratio]{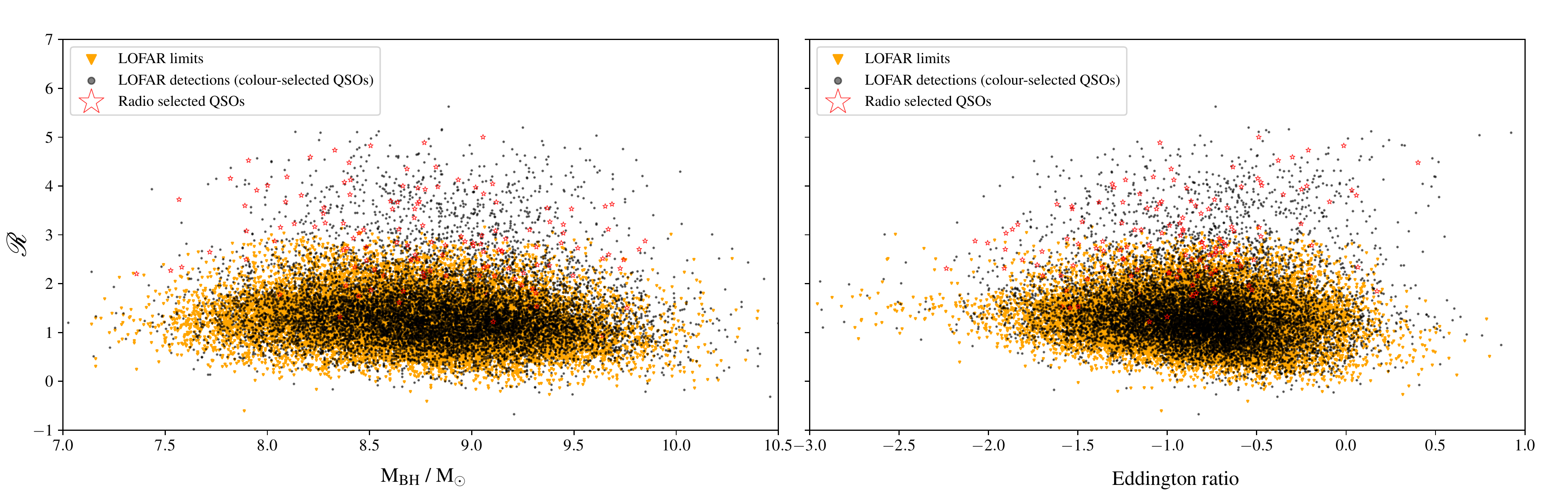}\\
\end{tabular}}
\caption{Left: Distribution of $\rr$ as a function of black-hole mass for LOFAR detections, limits of optically selected quasars, and radio selected quasars. Right: The distribution of $\rr$ as a function of Eddington ratio for LOFAR detections, limits of optically selected quasars, and radio selected quasars. \label{R-mass-accretion}}
\end{center}
\end{figure*}

\begin{figure*}
\begin{center}
\scalebox{1.7}{
\begin{tabular}{c}
\includegraphics[width=11cm,height=11cm,angle=0,keepaspectratio]{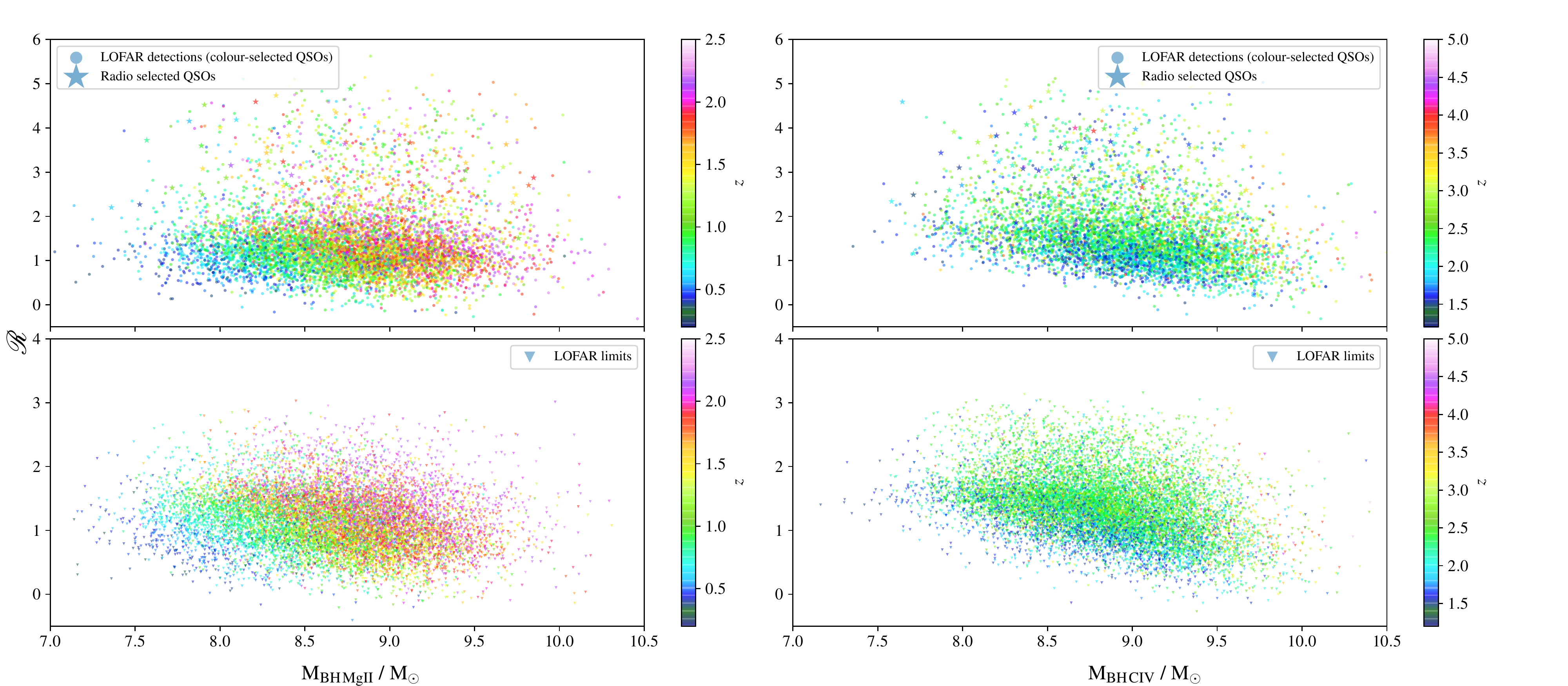}\\
\end{tabular}}
\caption{Top left and right: Distribution of $\rr$ as a function of black-hole mass measured using $\ion{}{MgII}$ and $\ion{}{CIV}$ lines, respectively, for LOFAR detections. Bottom left and right: The $\rr$ distribution of limits as a function of black-hole mass measured using $\ion{}{MgII}$ and $\ion{}{CIV}$ lines, respectively. \label{BHmasstest}}
\end{center}
\end{figure*}

We initially explored the relation between $\rr$ and black hole mass, and Eddington ratio ($\lambda_{\mathrm{Edd}}$) for all quasars (independent of whether they were detected by LOFAR). These are shown Fig. \ref{R-mass-accretion}. In the left panel we show the distribution of $\rr$ versus black hole mass and in the right panel that of $\lambda_{\mathrm{Edd}}$. In both panels we see a similar picture: $\rr$ does seem to weakly change with increasing black hole mass or $\lambda_{\mathrm{Edd}}$. This is true for both detections, limits, and radio selected objects. In order to quantitatively evaluate for correlations between $\rr$ and quasar nuclear properties we calculated Spearman's correlation coefficient and $p-$values. Results of the correlation analyses are given in Table \ref{spearman}. These results suggest that there is a weak negative but significant correlation between $\rr$ and black hole mass for both optically and radio selected quasars. There is also a weak but significant anti-correlation between $\rr$ and Eddington ratio for optically selected quasars. We cannot reject the null hypothesis for radio selected quasars: $\rr$ and Eddington ratio are uncorrelated. We investigate the downward trend of $\rr$ with increasing black hole mass further by probing the relation between $\rr$ and black hole masses measured using $\ion{}{MgII}$ line and those using $\ion{}{CIV}$ line. These can be seen in Fig. \ref{BHmasstest}. The downward trend of $\rr$ with black hole mass estimated using $\ion{}{CIV}$ line (for both detections and limits) is more apparent in the right panels of Fig. \ref{BHmasstest}. This might be intrinsic or due to decrease in signal-to-noise (S/N) of $\ion{}{CIV}$ line with increasing redshift \citep{Shen11}.

\begin{table*}
\caption{Spearman’s correlation and partial correlation coefficients between quasar properties and $\rr$. \label{spearman}}
\begin{center}
\begin{tabular}{lrrrr}
\hline
Sample category & Quasar nuclear property &Spearman’s correlation coefficients&Partial correlation\\ 
\hline
Colour selected and detected in LOFAR & black hole Mass and $\rr$ & -0.10, $p<0.0001$&-0.24, $p<0.0001$\\
&Eddington ratio and $\rr$&-0.14, $p<0.0001$&-0.22, $p<0.0001$\\
&$L_{\mathrm{i\,band}}$ and $\rr$&&-0.48, $p<0.0001$\\
Radio selected and detected in LOFAR& black hole Mass and $\rr$ & -0.27, $p=0.002$&-0.29, $p=0.001$\\
&Eddington ratio and $\rr$ & 0.07, $p=0.56$&0.05, $p=0.56$\\
&$L_{\mathrm{i\,band}}$ and $\rr$&&-0.43, $p<0.0001$\\
All quasars detected in LOFAR & black hole Mass and $\rr$ & -0.16, $p<0.0001$&-0.24, $p<0.0001$\\
&Eddington ratio and $\rr$ & -0.12, $p<0.0001$&-0.23, $p<0.0001$\\
&$L_{\mathrm{i\,band}}$ and $\rr$&&-0.48, $p<0.0001$\\
\hline
\end{tabular}
\end{center}
\end{table*}

\begin{figure*}
\begin{center}
\scalebox{2.1}{
\begin{tabular}{c}
\hspace{-3em}\includegraphics[width=11cm,height=11cm,angle=90,keepaspectratio]{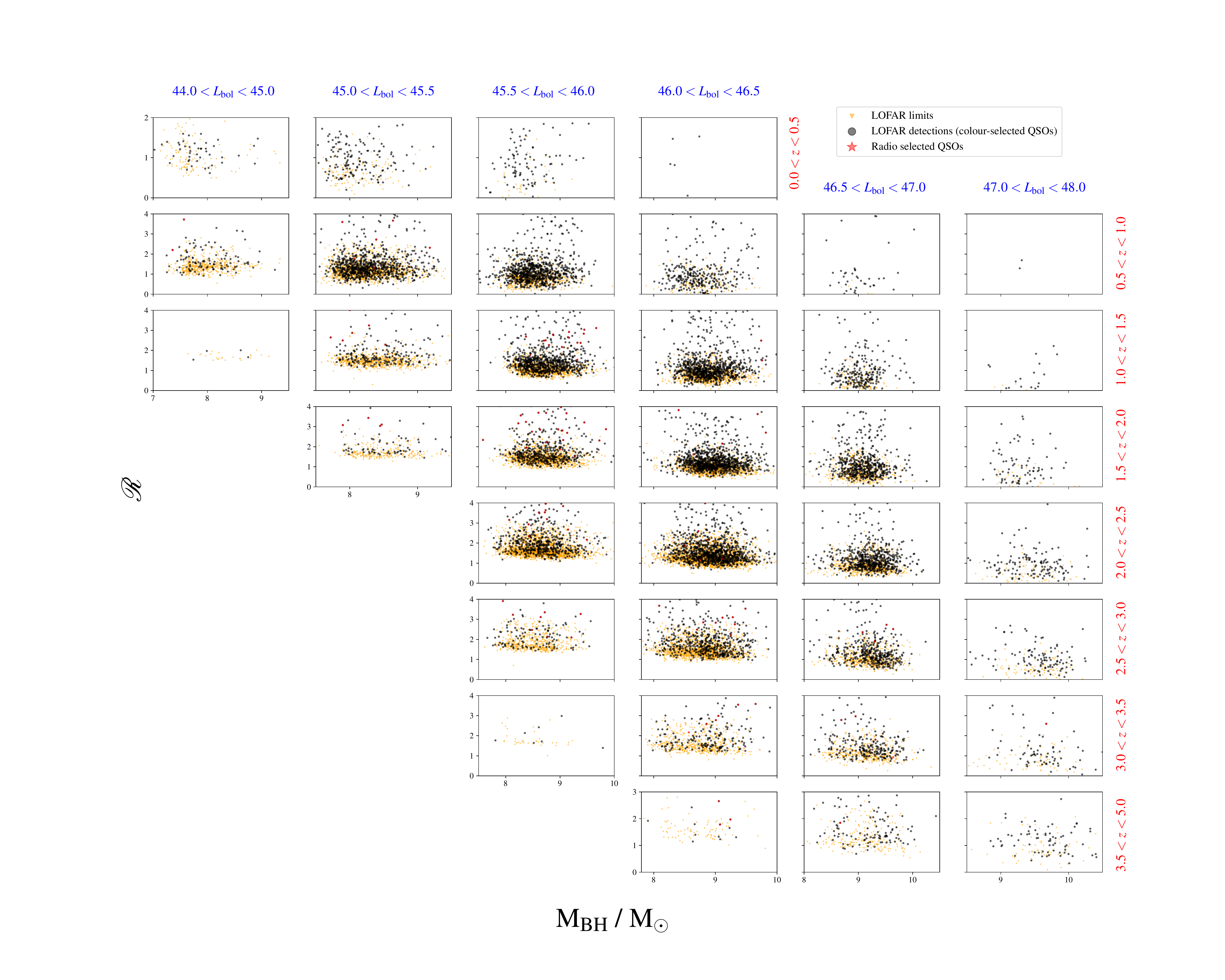}\\
\end{tabular}}
\caption{Distribution of $\rr$ as a function of black-hole mass for detections, limits, and radio selected quasars. Grey points are LOFAR detections, orange points are limits, and red points radio selected quasars. \label{match-mass}}
\end{center}
\end{figure*}

\begin{figure*}
\begin{center}
\scalebox{2.1}{
\begin{tabular}{c}
\hspace{-3em}\includegraphics[width=11cm,height=11cm,angle=90,keepaspectratio]{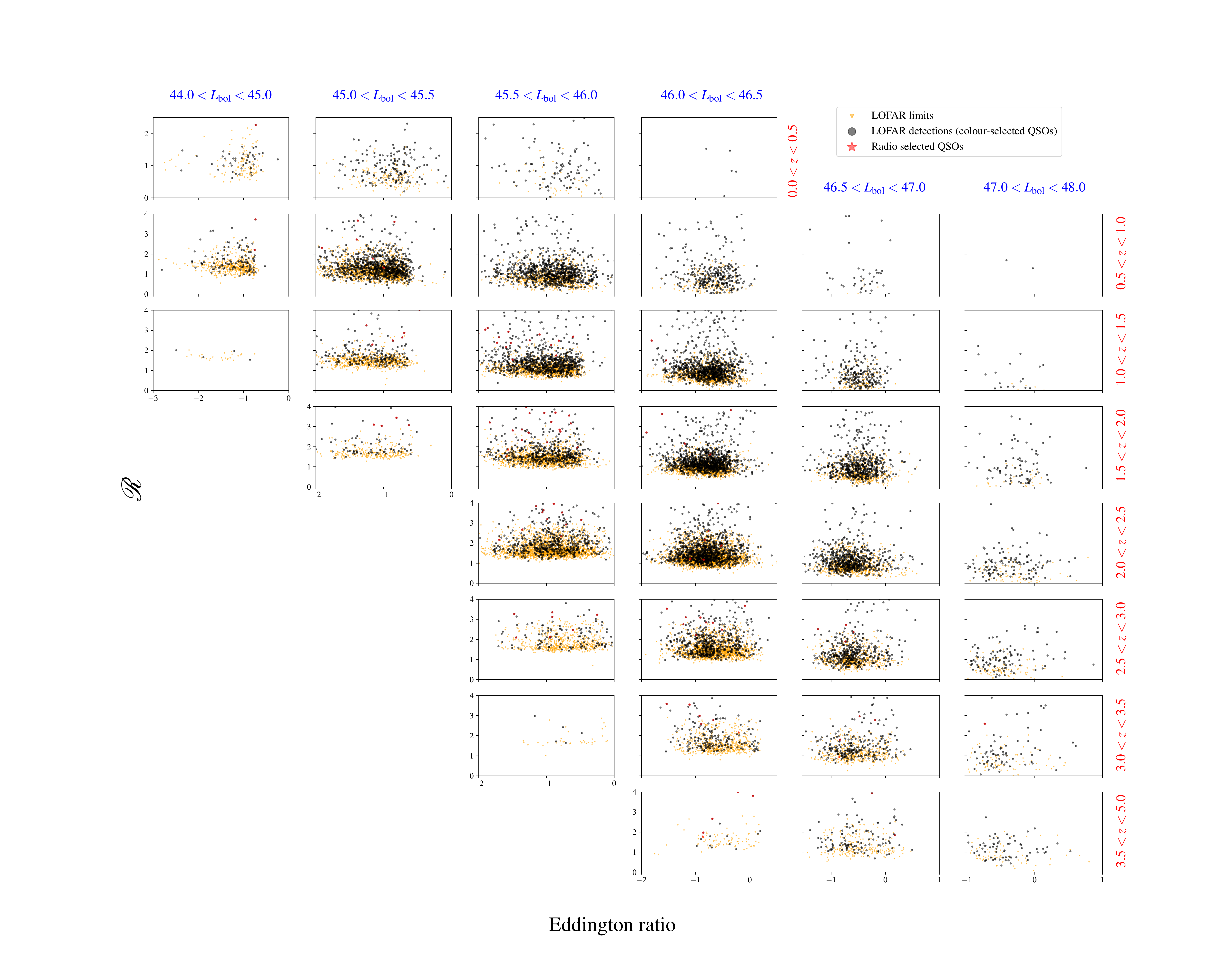}\\
\end{tabular}}
\caption{Distribution of $\rr$ as a function of Eddington ratio for detections, limits, and radio selected quasars. Grey points are LOFAR detections, orange points are limits, and red points radio selected quasars.\label{match-edd}}
\end{center}
\end{figure*}

Our sample size is very large and covers a wide redshift range so it is crucial to investigate relations for a sample matched in $z$ (and if possible in different galaxy parameters). Therefore, we match the sample in $L_\mathrm{{bol}}$ and $z$ and investigate the relation between $\rr$ and quasar nuclear properties. These can be seen in Fig. \ref{match-mass} and \ref{match-edd}. The lack of clear dependence on mass or accretion rate is still valid: there is a weak anti-correlation between $\rr$ and black hole mass or $\lambda_{\mathrm{Edd}}$. We performed partial correlation analysis for both optically and radio selected quasars to see if there is a correlation between $\rr$ and the nuclear parameters of quasars while controlling for the effect of redshift. Results of this analysis are similar to Spearman correlation test: there is a weak but significant anti-correlation between $\rr$ and black hole mass for both optically and radio selected sources. Although we see a weak correlation between $\rr$ and Eddington ratio for optically selected sources this is not clear for radio selected objects. These results are also given in Table \ref{spearman}.

\section{Discussion}

\label{sec:discuss}

We have used a sample of optically selected quasars from SDSS and investigated their low-frequency radio properties using LOFAR data over the HETDEX region, which was surveyed as part of the LOTSS at an average frequency of 144 MHz (Shimwell et al. 2018), as well as over the H-ATLAS/NGP field \citep{Hardcastle+16}. Taking into account the sample size, the combination of sensitivity and frequency provided by LOFAR has allowed us to make one of the most complete studies of individual quasars to date. The analyses presented in this study make use of visual inspection of a fraction (12$\%$) of these radio sources. We provide a summary of results derived from this work and compare with the literature by addressing the following questions:

\begin{itemize}

\item{\textbf{Is there a radio loudness dichotomy in quasars?}}\\

Figs. \ref{loudness} and \ref{loudness-z} clearly show that for a given $i$-band luminosity (i.e. a proxy for accretion luminosity) quasars have a wide range of radio luminosities. In other words, the $\rr$ distributions in both figures do not show any clear bimodality and this is also valid for different redshift bins. An evaluation of the odds ratio for a single Gaussian model against a Gaussian mixture model with two components suggests that either model is not favoured by the data.
In the light of these results we can conclude that there is no clear evidence for bimodality in the population. This is somewhat similar to what is observed for lower luminosity sources \citep[e.g.][]{2014ref50,Mingo16}. It is worth noting that this does not imply that there is only one mechanism powering the radio emission. This conclusion simply implies that if there is more than one mechanism, there is a smooth transition between the dominant mechanism as a function of $\rr$; there should be a number of sources in which the radio continuum might well be a combination of radio emission from small-scale jets as well as star formation.

As pointed out in the introduction, in the literature we see a range
of conclusions based on evaluation of quasars using the radio loudness
parameter. Various studies have found a uniform distribution of $\rr$ 
\citep[e.g.][]{Falcke+96,White+00,Lacy+01,Brotherton+01,Cirasuolo+03,Cirasuolo+032,Miller+11,Balokovic+12,Ballo+12} while other works have suggested that there is a bimodal $\rr$ distribution
\citep[i.e. there are two distinct quasar populations; e.g.][]{Ivezic+02,White+07}. Our results in the present paper are consistent with the idea that there is a wide continuum of radio properties in quasars for a given accretion luminosity.

Why does this disagreement in the literature persist? Differences in the methods used in past studies, including selection effects, affect the conclusions drawn. These are
the following:
\begin{enumerate}[i)]
  \item The classification ratios defined to date are not consistent; a source can be classified as RL quasar according to one classification and RQ quasar for another.
  \item The definition of radio loudness involves using fluxes (or luminosities) at various optical and radio bands \citep[e.g.][]{Kellermann+89,Falcke+96,Stocke+92,Ivezic+02,Jiang+07}.
  \item The construction of the radio loudness definitions to date have been based on samples from different surveys and samples with varying properties (such as size, redshift etc.).
\end{enumerate}

Finally, as pointed out by \cite{Miller+11} with the deeper data sets we are able to fill in the gaps between radio bright and radio faint objects, interpreted as a dichotomy in the literature. Our results support this argument.

In this work we address the above points in the best possible way. We start with a large sample of quasars and make use of low-frequency radio observations in order not to be dominated by Doppler enhancement. Unprecedented sensitivity of the LOFAR data enables us to detect a considerable percentage ($\sim 50\%$) of quasars. We avoid any selection bias by probing quasars selected differently (i.e. optically or radio selected) and include limits in most of our analyses. We particularly avoid classifying sources as RL or RQ using traditional ratios, and instead we assess the relation between $\rr$ and several quasar properties to reach solid conclusions.\\

\item{\textbf{Does the radio loudness depend on nuclear properties?}}\\

We showed that the radio loudness parameter estimated using $L_{144}$ and $L_{\mathrm{i\,band}}$ does not strongly depend on either black hole mass or $\lambda_{\mathrm{Edd}}$ for both optically and radio selected quasars (Fig. \ref{R-mass-accretion}). This is also true when we match the quasar sample in redshift and $L_{\mathrm{bol}}$. We investigated the same relations for LOFAR limits and obtained similar results (Fig. \ref{match-mass} and \ref{match-edd}). 

These relations have been investigated before using samples (mostly small in size) and data sets at different wavelengths. For instance \cite{2000Laor} found a correlation between black hole mass and the radio loudness parameters of AGN: quasars with M$_{\mathrm{black hole}} <3\times10^{8}$ M$_{\odot}$ are practically all RQt whereas nearly all PG quasars with M$_{\mathrm{black hole}}>3\times10^{9}$ M$_{\odot}$ are RL. \cite{mclure04} investigated optically selected quasars and found that radio bright quasars harbour black hole masses that are typically 0.16 dex (45 per
cent) more massive than those of their RQ counterparts. They also reported a strong correlation between radio loudness parameter and black hole mass when they combined radio bright and radio faint quasars. \cite{2006Metcalf} found that RL quasars on average have higher black holes masses than RQ quasars. \cite{2010Shankar} and \cite{2002Ho} did not observe any dependence of the radio loudness parameter on black hole mass whereas our results suggest that there is a weak anti-correlation between $\rr$ and black hole mass of optically selected quasars.

With regard to the relation between $\rr$ and $\lambda_{\mathrm{Edd}}$ a number of studies have found an inverse correlation between these two quantities: as $\lambda_{\mathrm{Edd}}$ increases $\rr$ decreases \citep[e.g.][]{2002Ho,Merloni+03,Nagar+05,Sikora+07}. This has been interpreted by \citet{2002Ho} as the switch between the accretion modes (i.e. radiatively efficient and radiatively inefficient) although such studies have often confused jet-related nuclear emission, which is present in both classes of object, with accretion-related nuclear emission, which is expected only in radiatively efficient sources \citep{Hardcastle+09}. However, by selection, a sample of quasars contains only radiatively efficient objects and therefore no such effect is expected. \citet{Sikora+07} suggested that RL quasars and RQ quasars show the same correlation but with a different normalisation (RL quasars to have higher $\rr$ values than RQ quasars); we see a weak anti-correlation in our data with $\rr$ (whether they have high $\rr$ or not). Recently, \citet{Ballo+12} analysed a sample of X-ray selected type 1 AGN and quasars and found that the radio loudness parameter is positively correlated with $\lambda_{\mathrm{Edd}}$. Similar to our findings \cite{2010Shankar} did not observe any dependency of $\rr$ on $\lambda_{\mathrm{Edd}}$. As discussed above, we must invoke disagreements about the sample definition and the definition of the radio loudness parameter in order to explain the contradictory results seen in the literature with regard to the $\rr-M_{\mathrm{black hole}}$ and $\rr-\lambda_{\mathrm{Edd}}$ relations.\\

\item{\textbf{What is the source of radio emission in quasars?}}\\

As described in Section \ref{sec:firdata} we have far-IR measurements over the H-ATLAS/NGP field. This has allowed us to evaluate the distribution of quasars in the far-IR to low-frequency radio luminosity plane incorporating the radio loudness information. Recently, \cite{Gurkan+18} investigated the low-frequency radio luminosity to star formation rate relation and far-IR to radio correlation (FIRC) in local star-forming galaxies selected based on their optical emission lines, using LOFAR-144 MHz measurements to probe radio and Herschel-250 $\mu$m to probe far-IR. In Fig. \ref{firc} we show the distribution of $L_{144}$ as a function of $L_{250}$ for the optically selected quasars. The black solid line shows the $L_{144} - L_{250}$ relation given by \cite{Gurkan+18}. The FIRC might be evolving with redshift, although it is not expected to be strong relative to its scatter \citep[e.g.][]{Calistro17,2009lf23}. Our quasar sample spans a wide redshift range ($0<z<5$) so we constrained the sample to $z<3.0$; there still might be a redshift evolution of FIRC for $z<3,$ so we indicate the expected flatter slope due to this evolution with a black arrow in Fig. \ref{firc}. 

Quasars with $\rr>2$ are above the FIRC, including $Herschel$ limits. Most quasars detected in both bands having $2<\rr<3$ are (on or) above the FIRC and quasars with $-1<\rr<2$ follow the FIRC. Quasars with $2.<\rr<6.$ have much higher radio luminosity for a given far-IR luminosity (including far-IR limits). It might be possible that low-frequency radio emission from these quasars (having $\rr \la 1$) are affected by star formation processes, although we cannot rule out small-scale jets producing such level of radio continuum emission (or combination of both). Follow-up high-resolution observations of selected objects will be invaluable for revealing the source of radio emission in these sources.

Evaluation of this figure suggests the following points: (i) not all quasars have significantly higher radio luminosities for a given far-IR luminosity, (ii) the fraction of quasars having higher radio luminosity increases with increasing $\rr$ as expected, (iii) quasars with $2<\rr<6$ tend to have higher radio luminosities than would be consistent with host-galaxy star formation assuming the FIRC. 

\begin{figure}

\begin{center}

\scalebox{0.8}{

\begin{tabular}{c}

\hspace{-4em}\includegraphics[width=13cm,height=13cm,angle=0,keepaspectratio]{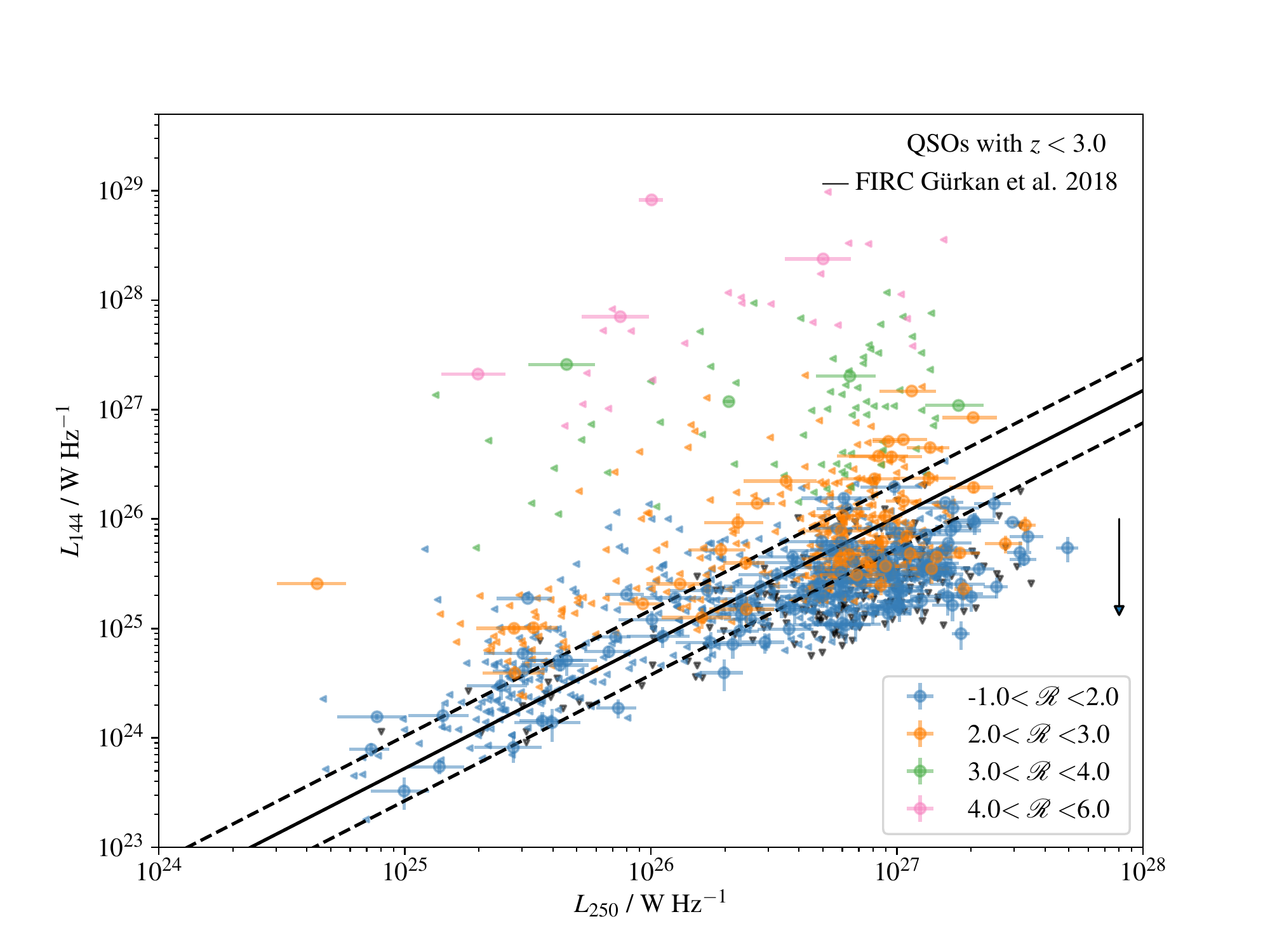}\\

\end{tabular}}

\caption{Distribution of $L_{144}$ of quasars with $z<3.0$ as a function of $L_{250}$. The solid black line shows the FIRC relation given by \citet{Gurkan+18}. Circles indicate sources detected in both bands: left-pointing arrows and black down-pointing arrows represent limits in 250 $\mu$m and 144-MHz bands. Points (except LOFAR limits) are colour coded by their $\rr$ values. We expect to get a flatter slope with increasing redshift due to the evolution of FIRC \citep{Calistro17}, which is indicated by a black arrow. \label{firc}}

\end{center}

\end{figure}

We further investigated whether the $\rr \sim 1.6$ peak we observe is driven by star formation using the AGN--SFR relation given by \cite{2015lf17} and the SFR--$L_{144}$ relation recently provided by \cite{Gurkan+18}. Using the relation given in the top panel of Fig. 9 in \cite{2015lf17} for RQ AGN (SFR $\propto$ $P_{\mathrm{AGN}}^{3.5}$) and the bolometric correction given by \cite[][the relation 11]{2012ref153} for an average $L_{i}$ star formation rate is expected to be around 12 M$_{\odot}$ yr$^{-1}$, which corresponds to $L_{144} \approx1.6\times10^{23}$ W Hz$^{-1}$. Then, the expected $\rr$ is $\sim1.5$. This estimated $\rr$ value, which is expected to be due to star formation, agrees well with the peak value we obtain for the $\rr$ of quasars. This agreement (within possible uncertainties) reinforces the idea that the radio emission in low luminosity of quasars might be mainly due to star formation, assuming that these objects follow the correlations observed in other samples between radio emission and star formation and star formation and AGN activity. On the other hand, higher resolution radio observations, which would allow us to separate SF and AGN, would provide a clear picture on this.


The $\rr$ distribution and the evaluations provided above indicate that at high radio luminosities the radio luminosity is most likely dominated by jet-related emission from these quasars. For a range of sources the radio luminosity will be presumably dominated by AGN, although we are not able to resolve these in all cases; high-resolution observations would be required to see jets at small scales. Jets are capable of producing radio emission at a very wide range of luminosities; the radio luminosity depends on the jet power, environment, and source age \citep{Hardcastle18}. However, the emission from low-power jets may be swamped by that from star formation, giving rise to the observed distribution of $\rr$ (Fig.\ \ref{loudness}).

\end{itemize}



\section{Conclusions}
\label{sec:discuss2}

As mentioned above earlier studies of quasars (and of radio galaxies) were constrained by the limitations of the available radio surveys, such as FIRST and NVSS, although they nevertheless provided important insights into quasar properties. The current innovations in the management of big data sets and in radio data reduction techniques have enabled us to carry out radio surveys at low radio frequencies with unprecedented angular resolution and sensitivity in relatively short observing times. In this paper we have compiled and analysed the largest quasar sample to date detected at low radio frequencies, which has provided a more complete picture of quasars in terms of the radio loudness parameter and its relation to the nuclear properties of quasars.

In the picture we favour in this work, AGN jets and star formation-related radio emission  can both operate in quasars and there is no RL/RQ dichotomy, but rather a smooth transition (probably with increasing jet power) between the dominance of the two processes.\footnote{In the cases in which SF and AGN co-exist in galaxies, decoupling the competing SF and AGN requires sub-arcsec resolution.} This helps to explain why studies that search for compact AGN-related emission generally find it in many objects \citep[e.g.][]{White+17}, while large-scale surveys of integrated radio emission, such as that of the present work or of \cite{kimball+11}, conclude that star formation dominates the low-$\rr$ population. We have shown that radio does not appear to depend strongly on (estimated) accretion rate or black hole mass. A key question is therefore ``What are the other parameters that might play a role in generating collimated powerful radio jets?", or, equivalently,``Why do most quasars (or in general AGN) not present these radio jets?"

Active galactic nuclei bolometric luminosity is a function of accretion rate, which is a function of black hole mass, and radiative efficiency, which
is a function of black hole spin \citep[][]{frank02}. The radio luminosity of AGN is a function of AGN jet power , radiative losses, time, cosmic epoch, and finally AGN environment; the AGN\ jet power is, in turn, expected to be
a different function of black hole spin, black hole mass, accretion rate,
and magnetic flux. Thus black hole spin is a key parameter that might be an answer to the above questions. This has been extensively discussed in the literature with inconclusive results as it is challenging to obtain reliable estimates of black hole spin (the details of black hole spin measurements are beyond the scope of this paper so we do not discuss them in this work). A recent study by \citet{Reynolds13} has shown a relation between black hole spin and black hole mass for a small sample of sources. There seems to be no clear relation between these two quantities, contrary to expectations. Additionally, \citet{Sikora+07} introduced a modified spin paradigm in which massive sources (such as elliptical galaxies in which we mostly find powerful radio sources) host a spinning black hole as a result of mergers in their history, while  moderate mass objects (i.e. spiral galaxies) have slowly spinning black holes. The spin direction of the black hole accretion disc with respect to the black hole spin has also been proposed as relevant to this question: sources with powerful jets are expected to have retrograde systems (the black hole spin and accretion disc counter rotate), which would generate highly energetic jets, whereas sources with small jets are thought to have prograde systems \citep[e.g.][]{Garofalo+10,Ballo+12}.

It is worthwhile to note that because of various effects (see the text above) we still do not know the relation between AGN jet power and the radio luminosity \citep[e.g.][]{2014ref89,Hardcastle18}, which is the measurement we usually attain, and the jet power--radio luminosity relation for different radio populations, i.e. FRI and FRII this relation is expected to be different, however (see \citealt{Croston+18}).

Finally, in the context of the big picture, understanding the main physical processes that cause an active source to generate strong jets is crucial for evaluating the intrinsic role of AGN and AGN feedback in galaxy evolution. Therefore, obtaining large volume limited samples and deep radio surveys, which would potentially be less affected by biases associated with ability to detect sources, is vital. The desired data for the cutting-edge science questions will be obtained with the current low-frequency surveys (such as LoTSS), next generation surveys (such as Tier2 -- LoTSS), WEAVE-LOFAR \citep{smith+16}, WEAVE-QSO (expected to provide high S/N spectra of quasars with $z>2$), and next-generation telescopes such as the SKA which will reach much higher sensitivities at low radio frequencies.

\begin{acknowledgements}
GG acknowledges the CSIRO OCE Postdoctoral Fellowship. MJH, WLW acknowledges support from the UK Science and Technology Facilities Council [ST/M001008/1]. PNB and JS are grateful for support from the UK STFC via grant ST/M001229/1. LKM acknowledges support from Oxford Hintze Centre for Astrophysical Surveys, which is funded through generous support from the Hintze Family Charitable Foundation. This publication arises from research partly funded by the John Fell Oxford University Press (OUP) Research Fund. IP acknowledges support from INAF under PRIN SKA/CTA ‘FORECaST’. MJJ acknowledges support from Oxford Hinzte Centre for Astrophysical Surveys, which is funded through generous support from the Hintze Family Charitable Foundation. KJD acknowledges support from the ERC Advanced Investigator programme NewClusters 321271. RKC is grateful for support from the UK STFC. JHC acknowledges support from the Science and Technology Facilities Council (STFC) under grants ST/R00109X/1 and ST/R000794/1. SM acknowledges funding through the Irish Research Council New Foundations scheme and the Irish Research Council Postgraduate Scholarship scheme.

This paper is based on data obtained with the International LOFAR Telescope (ILT) under project codes LC2 038 and LC3 008. LOFAR \citep{vanHaarlem2013} is the LOw Frequency ARray designed and constructed by ASTRON. It has observing, data processing, and data storage facilities in several countries, which are owned by various parties (each with their own funding sources), and which are collectively operated by the ILT foundation under a joint scientific policy. The ILT resources have benefited from the following recent major funding sources: CNRS-INSU, Observatoire de Paris and Universit'e d’Orl'eans, France; BMBF, MIWF-NRW, MPG, Germany; Science Foundation Ireland (SFI), Department of Business, Enterprise and Innovation (DBEI), Ireland; NWO, The Netherlands; The Science and Technology Facilities Council, UK[7]. Part of this work was carried out on the Dutch national e-infrastructure with the support of SURF Cooperative through grant e-infra 160022 and we gratefully acknowledge support by N. Danezi (SURFsara) and C. Schrijvers (SURFsara).

This research has made use of the University of Hertfordshire high-performance computing facility (\url{http://stri-cluster.herts.ac.uk/}) and the LOFAR-UK computing facility located at the University of Hertfordshire and supported by STFC [ST/P000096/1]. This research made use of {\sc astropy}, a community-developed core Python package for astronomy \citep{2013lf47} hosted at \url{http://www.astropy.org/} and of {\sc topcat} \citep{2005lf46}. 

This publication uses data generated via the Zooniverse.org platform, development of which is
funded by generous support, including a Global Impact Award from Google, and by a grant from the Alfred P.
Sloan Foundation.

\textit{Herschel}-ATLAS is a project with \textit{Herschel}, which is an ESA space observatory with science instruments provided by European-led Principal Investigator consortia and with important participation from NASA. The H-ATLAS website is \url{http://www.H-ATLAS.org/}.

Funding for SDSS-III has been provided by the Alfred P. Sloan Foundation, the Participating Institutions, the National Science Foundation, and the U.S. Department of Energy Office of Science. The SDSS-III web site is \url{http://www.sdss3.org/}.

SDSS-III is managed by the Astrophysical Research Consortium for the Participating Institutions of the SDSS-III Collaboration including the University of Arizona, the Brazilian Participation Group, Brookhaven National Laboratory, Carnegie Mellon University, University of Florida, the French Participation Group, the German Participation Group, Harvard University, the Instituto de Astrofisica de Canarias, the Michigan State/Notre Dame/JINA Participation Group, Johns Hopkins University, Lawrence Berkeley National Laboratory, Max Planck Institute for Astrophysics, Max Planck Institute for Extraterrestrial Physics, New Mexico State University, New York University, Ohio State University, Pennsylvania State University, University of Portsmouth, Princeton University, the Spanish Participation Group, University of Tokyo, University of Utah, Vanderbilt University, University of Virginia, University of Washington, and Yale University.

The National Radio Astronomy Observatory (NRAO) is a facility of the National Science Foundation operated under cooperative agreement by Associated Universities, Inc.

\end{acknowledgements}

\bibliographystyle{aa} 
\bibliography{ref} 
%
%

\begin{appendix} %

\onecolumn
\begin{figure}
\begin{xtabular}{cccc}
\hspace{-1.em}\includegraphics[scale=0.15]{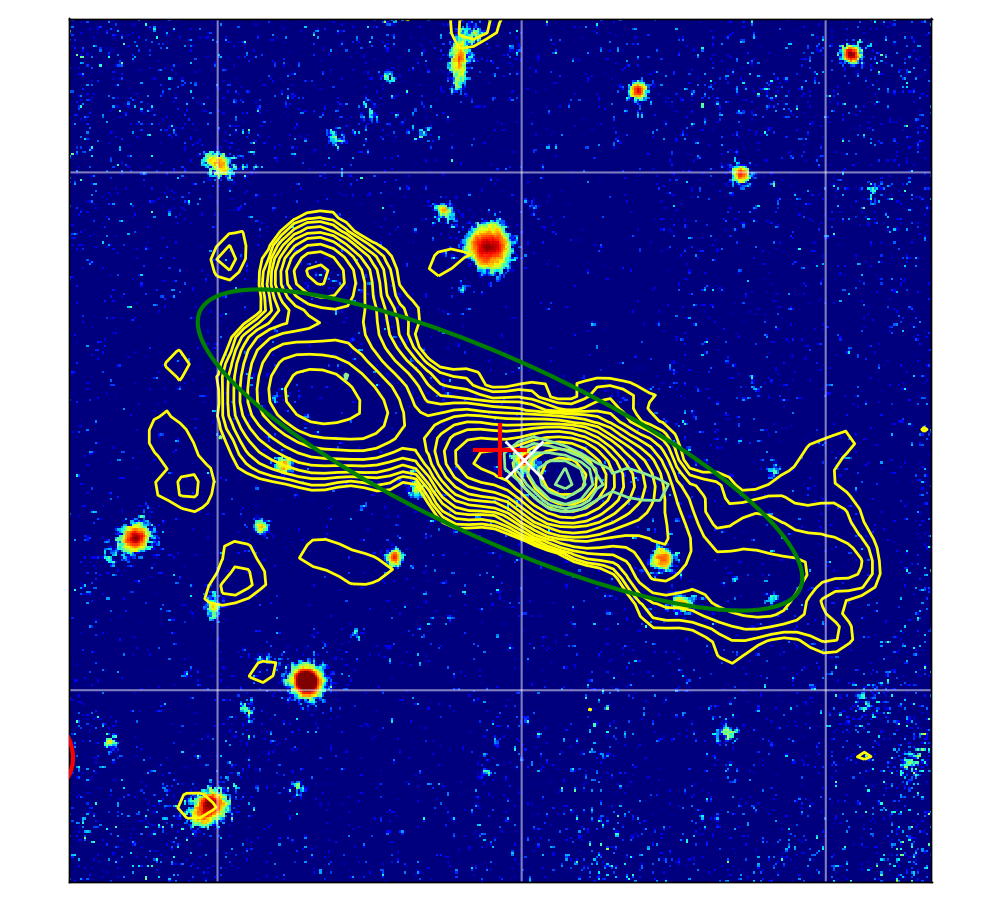}&
\hspace{2.em}\includegraphics[scale=0.15]{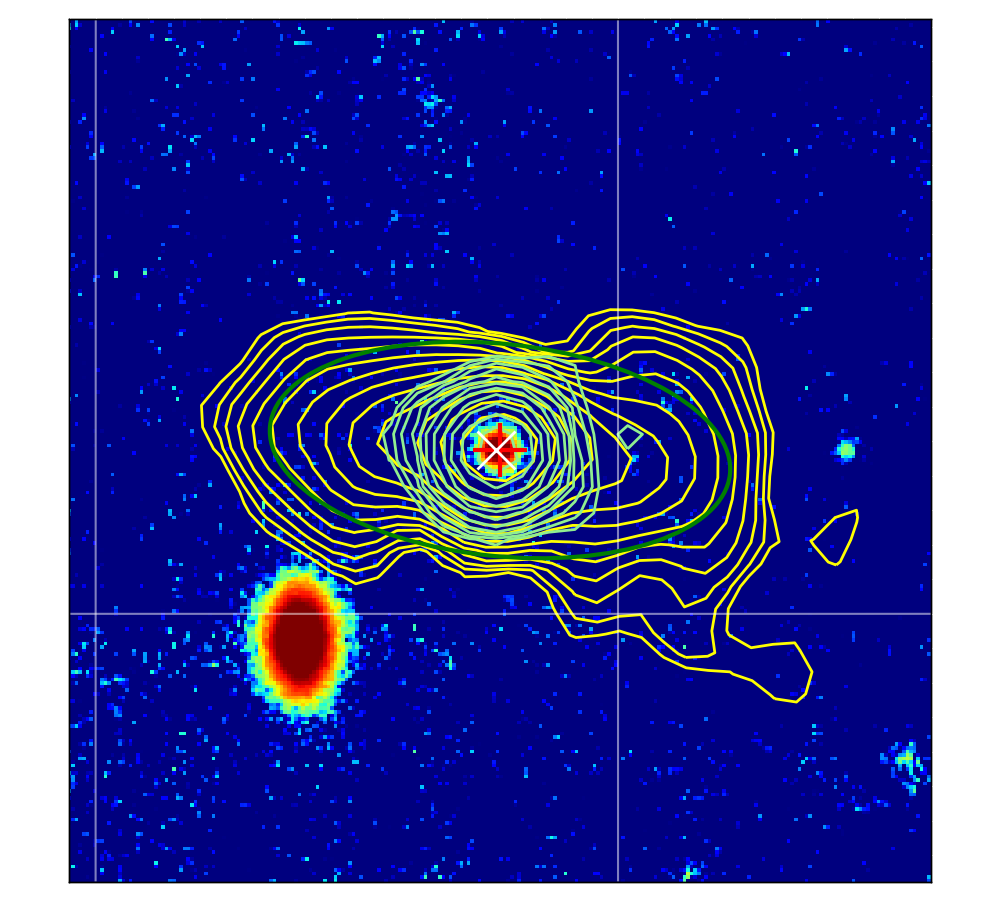}&
\hspace{2.em}\includegraphics[scale=0.15]{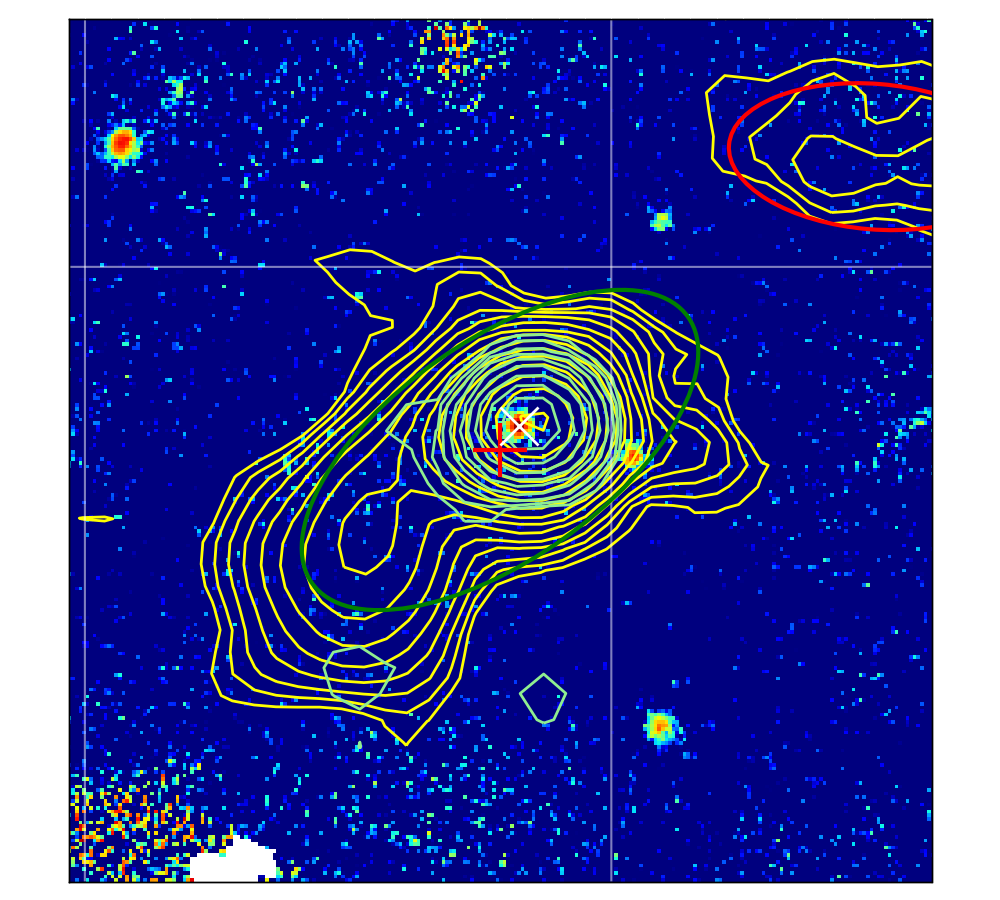}&
\hspace{2em}\includegraphics[scale=0.15]{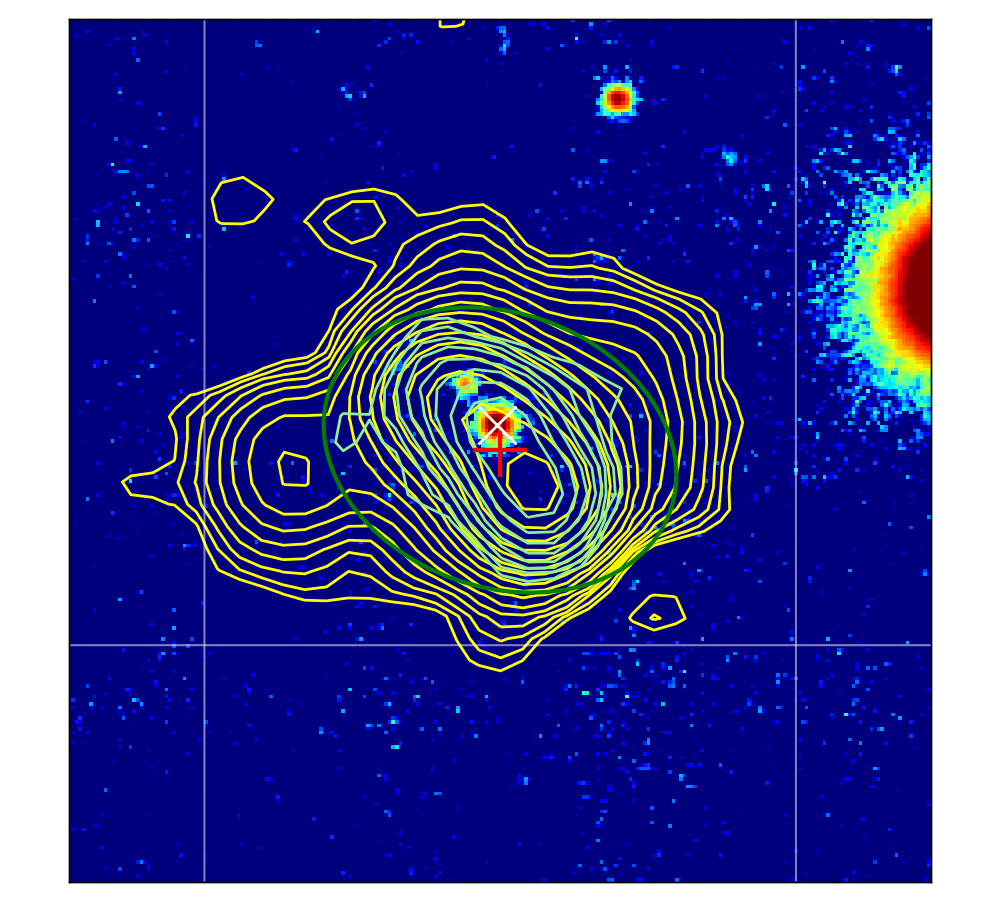}\\
\hspace{-1.em}\includegraphics[scale=0.15]{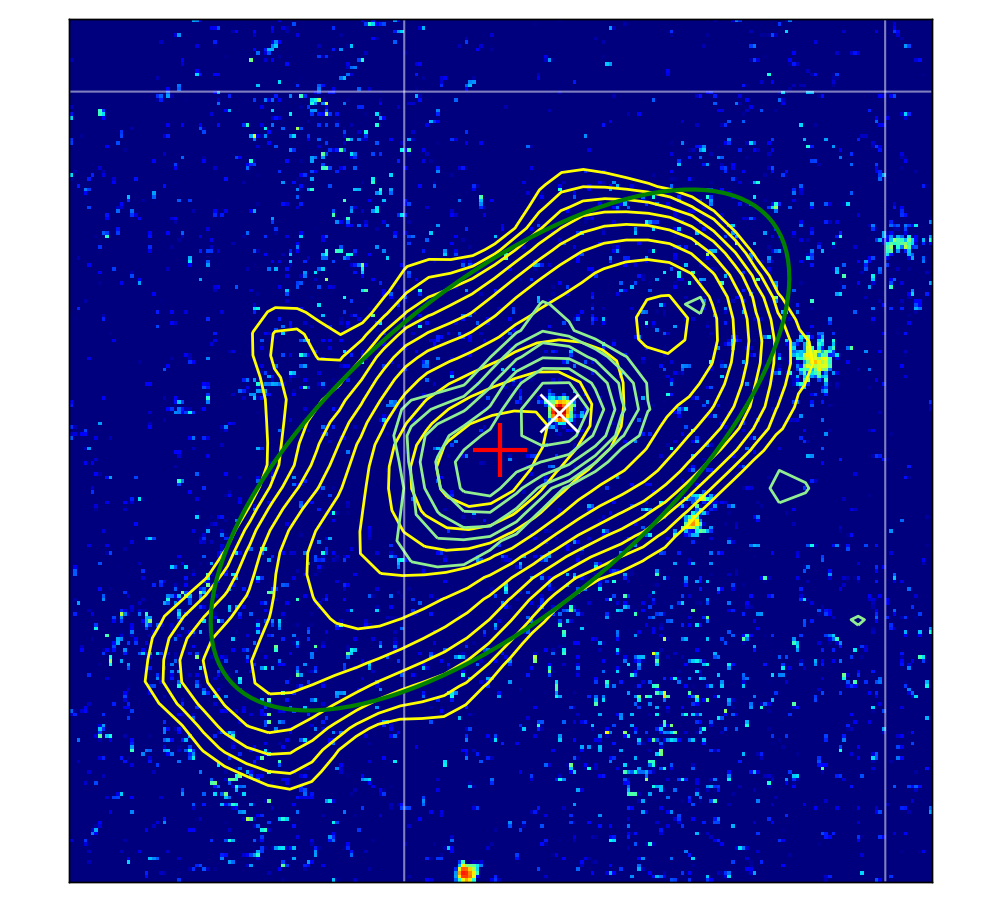}&
\hspace{2.em}\includegraphics[scale=0.15]{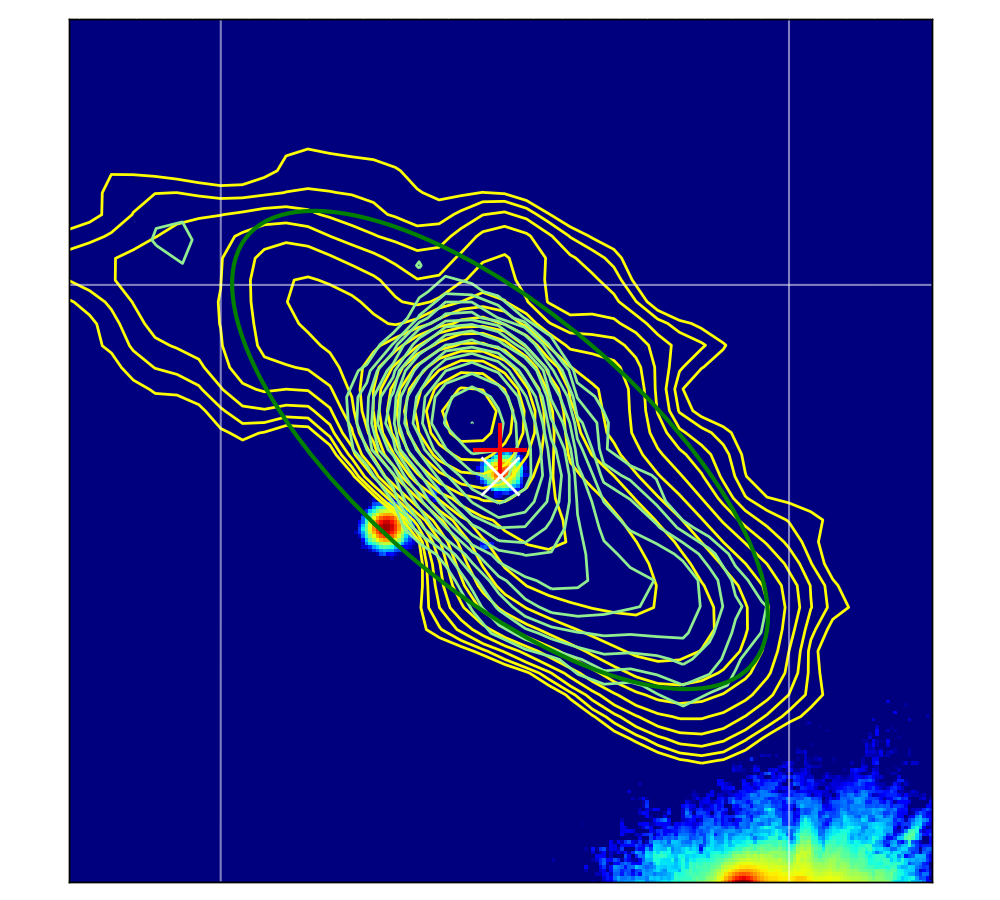}&
\hspace{2.em}\includegraphics[scale=0.15]{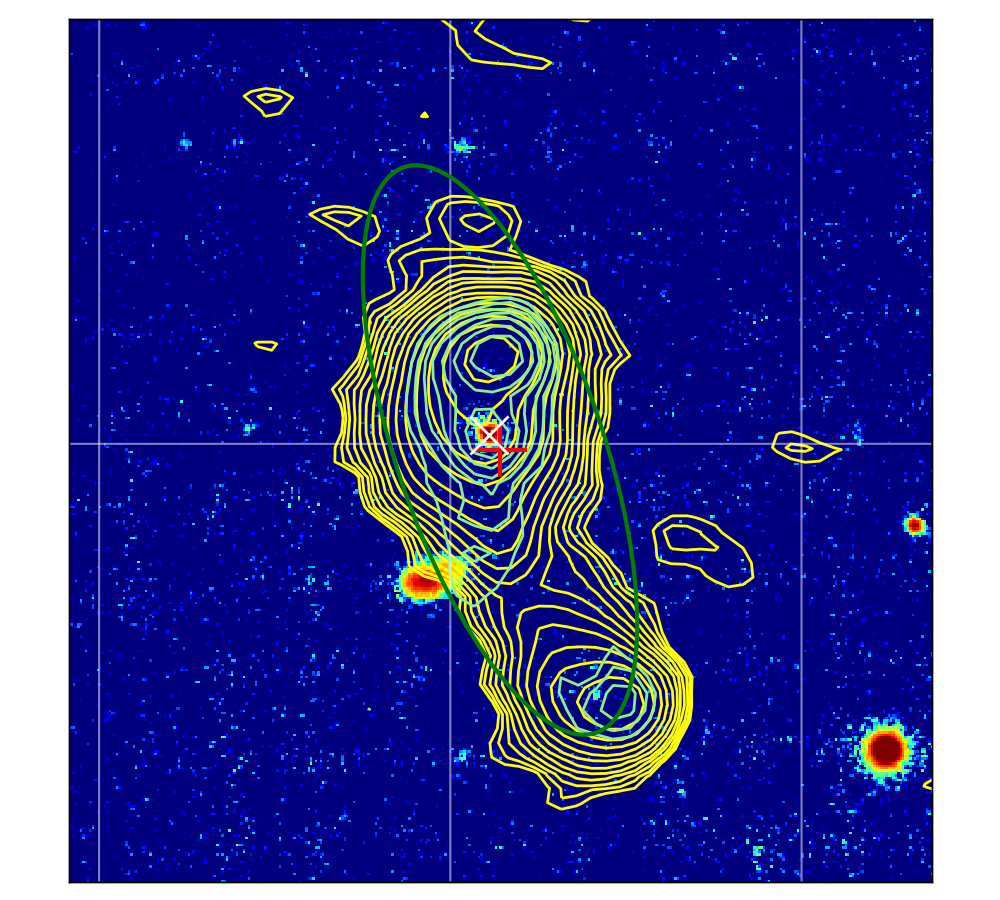}&
\hspace{2.em}\includegraphics[scale=0.15]{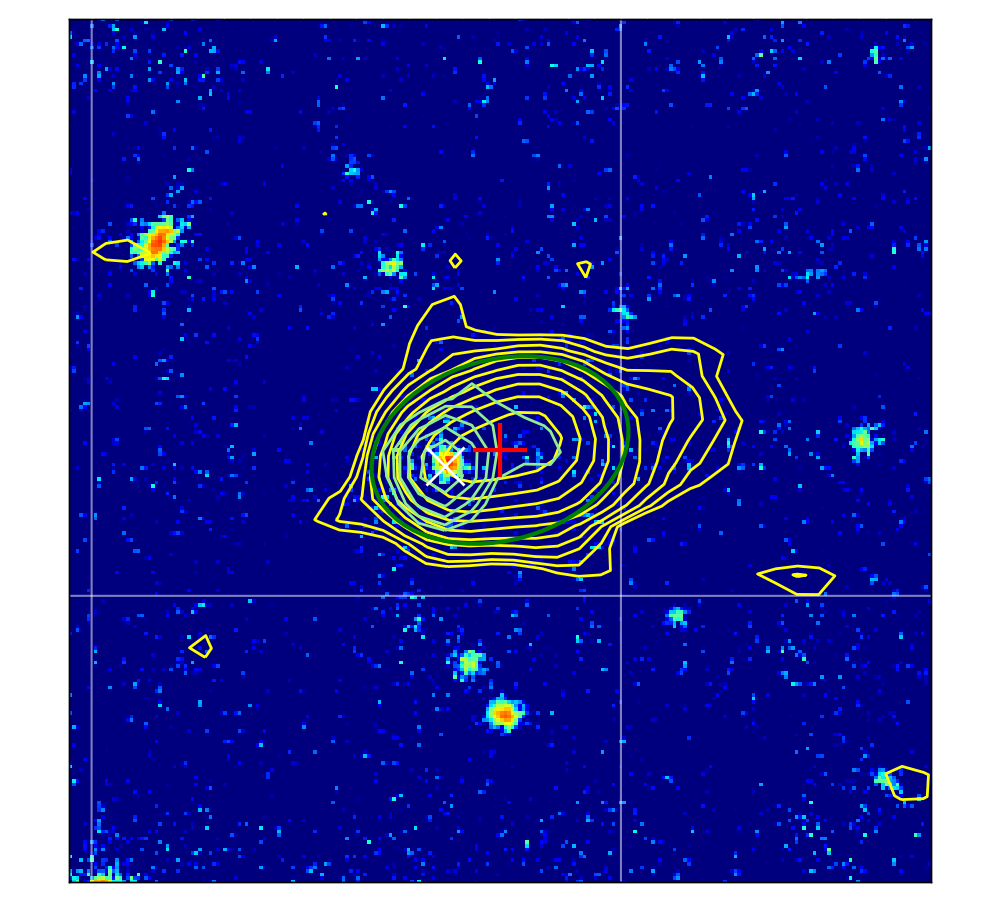}\\
\hspace{-1.em}\includegraphics[scale=0.15]{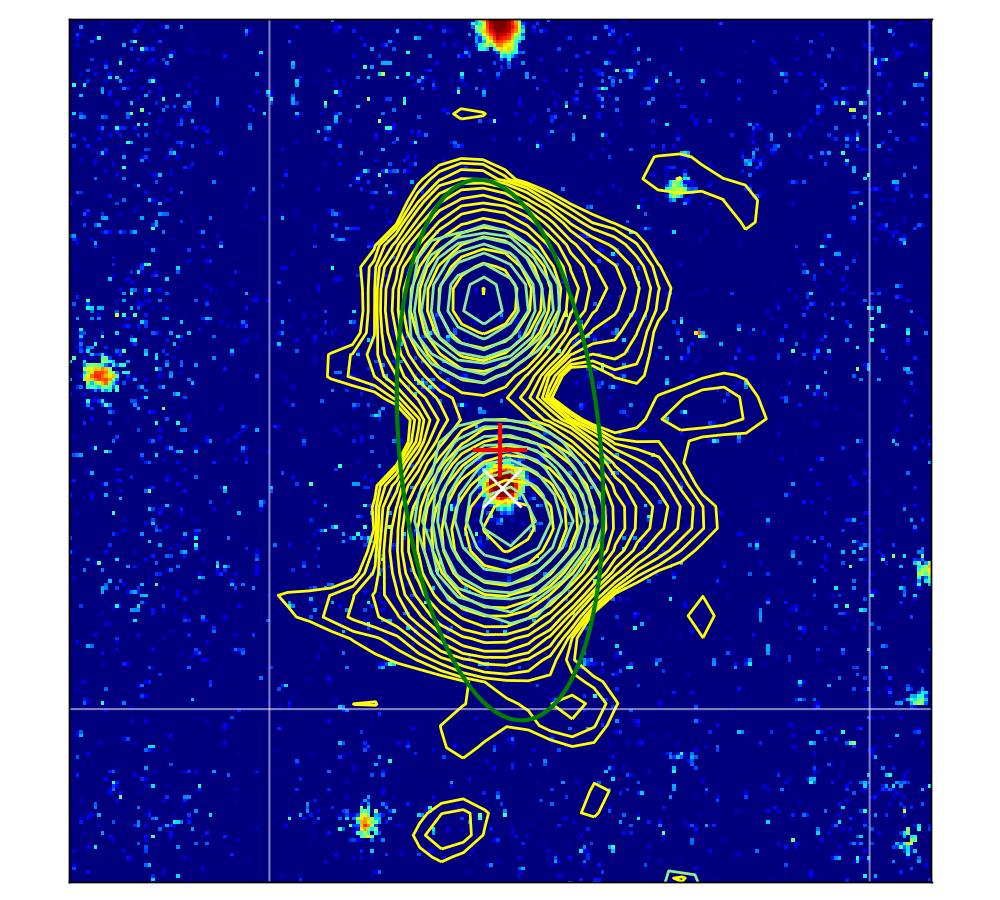}&
\hspace{2.em}\includegraphics[scale=0.15]{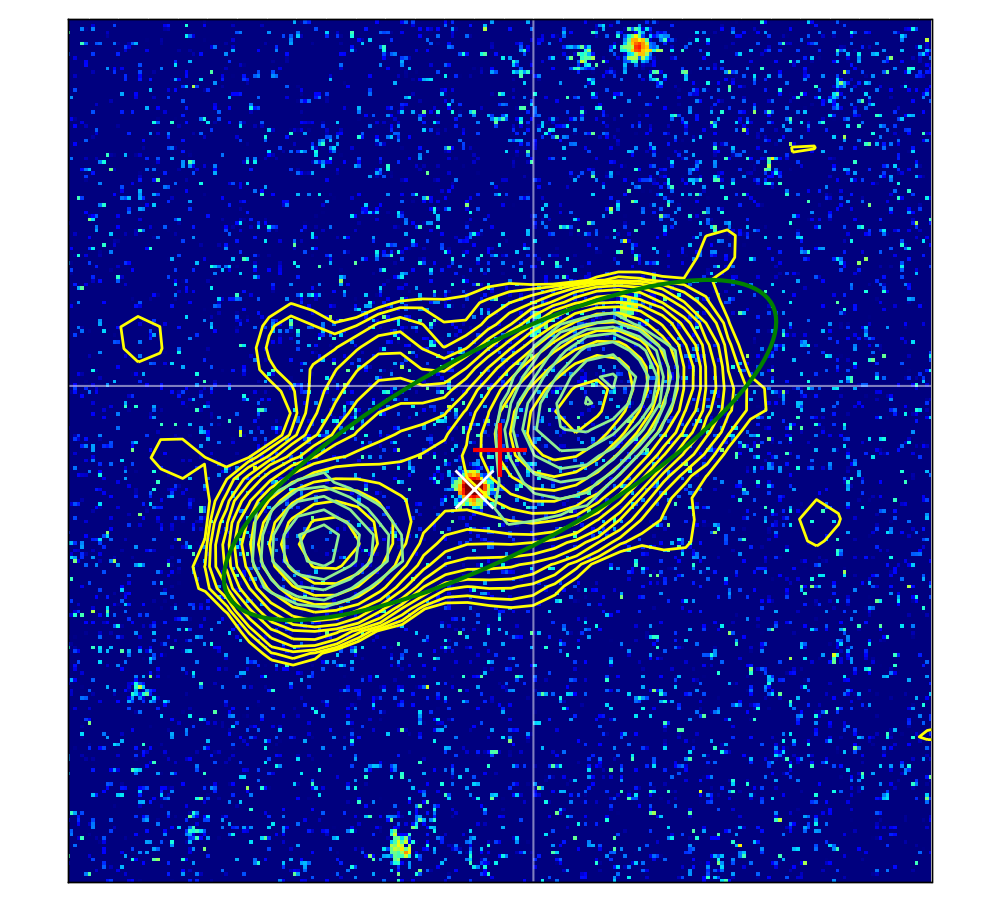}&
\hspace{2.em}\includegraphics[scale=0.15]{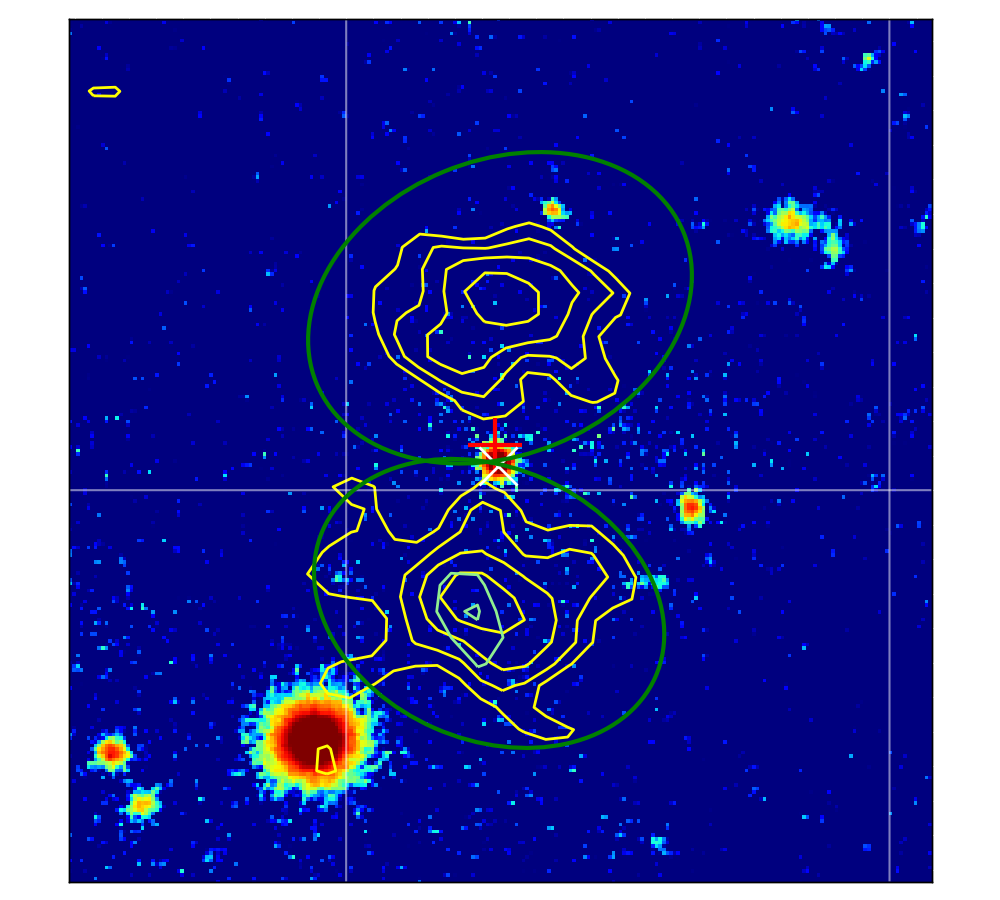}&
\hspace{2.em}\includegraphics[scale=0.15]{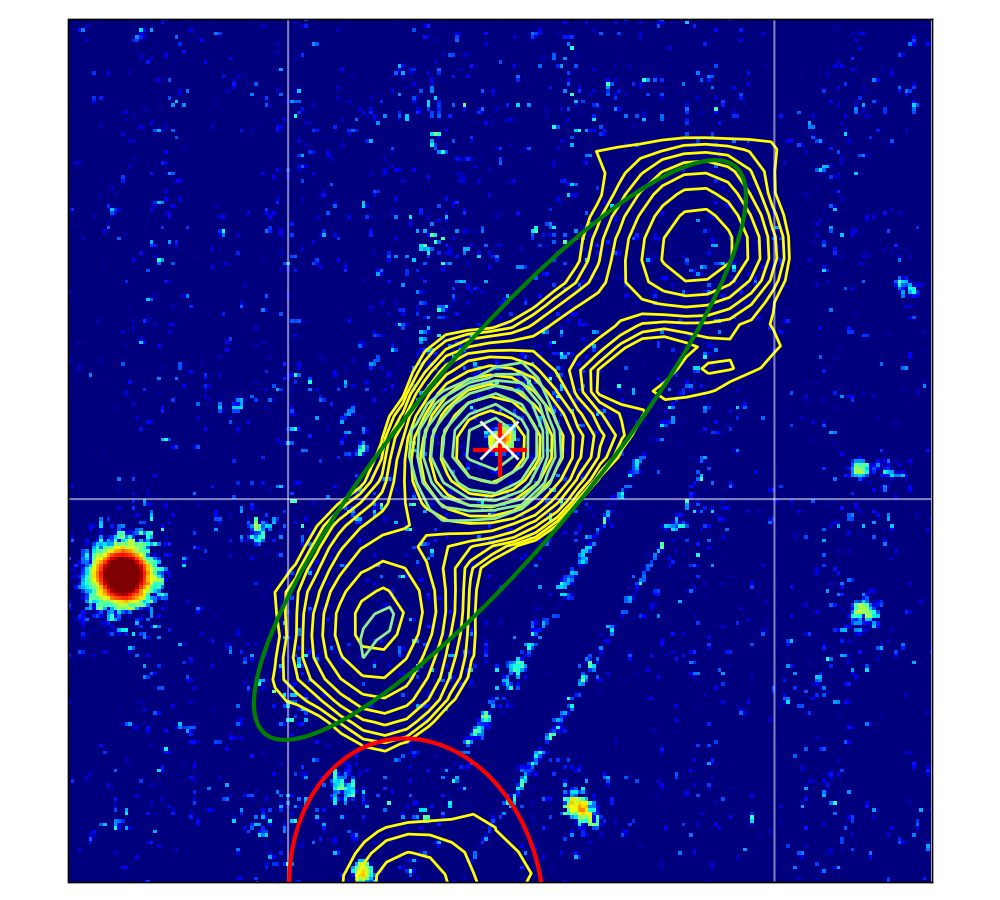}\\
\end{xtabular}
\caption{Examples of our extended sources identified by visual inspection: LOFAR contours (yellow) and FIRST contours (green) on PanSTARRS optical images (colour). Vertical grid spacing is 1 arcmin. \label{App}}
\end{figure}

\end{appendix}

\end{document}